\documentclass[draft]{agujournal2019}
\usepackage{url} 
\usepackage{lineno}
\usepackage[finalnew]{trackchanges} 
\usepackage{soul}
\usepackage[utf8]{inputenc}
\usepackage{amsmath}
\usepackage{amssymb}
\usepackage{graphicx}
\usepackage{subcaption}

\journalname{Earth's Future}

\begin{document}

%
%

\title{Forecasting the 2016-2017 Central Apennines Earthquake Sequence with a Neural Point Process}

%
%




\authors{Samuel Stockman \affil{1}, Daniel J. Lawson \affil{1}, Maximilian J. Werner \affil{2}}


\affiliation{1}{School of Mathematics, University of Bristol}
\affiliation{2}{School of Earth Sciences, University of Bristol}




\correspondingauthor{Samuel Stockman}{sam.stockman@bristol.ac.uk}




\begin{keypoints}
\item We construct a new machine learning variant of point processes for short-term earthquake forecasting enhanced catalogs.
\item The neural point process gains higher forecasting performance from the low magnitude data than ETAS and is faster to train.
\item This forecasting performance on the 2016 Central Italy sequence motivates continued development in this class of models.
\end{keypoints}

%
%

%
%


\begin{abstract}
Point processes have been dominant in modeling the evolution of seismicity for decades, with the Epidemic\change{ }{-}Type Aftershock Sequence (ETAS) model being most popular. Recent advances in machine learning have constructed highly flexible point process models using neural networks to improve upon existing parametric models. We investigate whether these flexible point process models can be applied to short-term seismicity forecasting by extending an existing temporal neural model to the magnitude domain and we show how this model can forecast earthquakes above a target magnitude threshold. We first demonstrate that the neural model can fit synthetic ETAS data, however, requiring less computational time because it is not dependent on the full history of the sequence. By artificially emulating short-term aftershock incompleteness in the synthetic dataset, we find that the neural model outperforms ETAS. Using a new enhanced catalog from the 2016-2017 Central Apennines earthquake sequence, we investigate the predictive skill of ETAS and the neural model with respect to the lowest input magnitude. Constructing multiple forecasting experiments using the Visso, Norcia and Campotosto earthquakes to partition training and testing data, we target M3+ events. We find both models perform similarly at previously explored thresholds (e.g., above M3), but lowering the threshold to M1.2 reduces the performance of ETAS unlike the neural model. We argue that some of these gains are due to the neural model's ability to handle incomplete data. The robustness to missing data and speed to train the neural model present it as an encouraging competitor in earthquake forecasting. 
\end{abstract}

\section*{Plain Language Summary}

For decades, the Epidemic-Type Aftershock Sequence (ETAS) model has been the most popular way of forecasting earthquakes over short time spans (days/weeks). It is formulated mathematically as a point process, a general class of statistical model describing the random occurrence of points in time. Recently the machine learning community have used neural networks to make point processes more expressive and titled them neural point processes. In this study we investigate whether a neural point process can compete with the ETAS model. We find that the two models perform similarly on computer simulated data; however, the neural model is much faster with large datasets and is not hindered if there is missing data for smaller earthquakes. Most earthquake catalogs contain missing data due to varying capability in our detection methods, therefore we need models that are robust to this missingness. We then find that the neural model outperforms ETAS on a new catalog for the 2016-2017 Central Apennines earthquake sequence, which through machine learning detection contains thousands of previously undetected small magnitude events. We argue that some of this improvement can in fact be explained by missing data. These results present neural point processes as an encouraging competitor in earthquake forecasting. 

\section{Introduction}
%

The construction of machine learning algorithms for detecting the arrival times of earthquake phases (eg. \citeA{zhu2019phasenet}) combined with an accelerated growth in the number of seismic sensors, has meant that sizes of earthquake catalogs have grown substantially \cite{kong2019machine}. With the amount of available seismicity data increasing, current forecasting methods that include the full history of the catalog \add{in the form of all event pairs }are increasingly inefficient and might not be flexible enough to incorporate this additional data, thus the need for the application of methods developed in the machine learning community is becoming more apparent. However, there exists a disconnect between the tools used by statistical seismologists and those in the machine learning community that apply their methods to seismic data. This work attempts to bridge that gap (to some extent) by considering a machine learning variant of point processes. Point processes are a class of models that contain the Epidemic\change{ }{-}Type Aftershock Sequence (ETAS) model, a widely accepted and used point process model for earthquakes \cite{ogata1988statistical,ogata1998space,marzocchi2014establishment,mancini2019improving}. In working with a machine learning method with a conditional intensity function (the function that explicitly defines a point process) \cite{zhuangtheme}, we present a model that is directly comparable to ETAS models, the current benchmark for short-term earthquake forecasting, yet now with desirable properties such as flexibility and scalability. The machine learning variant of point processes we introduce are known as neural point processes.\\
At the heart of neural point processes is the use of a recurrent neural network (RNN) to learn a compact representation of the history of events, first introduced by \citeA{du2016recurrent}. The sequential nature of the way data pass through RNNs makes them an ideal modeling tool for temporal data. Instead of directly summing over all past events, as in models based on the Hawkes process \cite{hawkes1971spectra}, a fixed length vector representation of the past is learnt and updated at each new time step. Forecasts can then be made by modelling the conditional intensity function on this vector  \cite{xiao2017modeling,li2018learning,upadhyay2018deep,huang2019recurrent,NEURIPS2019_39e4973b}, or instead through directly modelling the probability of the next event \cite{shchur2019intensity}. We direct the reader to \citeA{shchur2021neural} for a review of neural point process models. None of these models, however, are directly suitable for describing the temporal behaviour of seismicity, which includes a continuous mark space for the magnitude of each earthquake as well as the times. We require a model that is dependent on continuous marks as well as forecasts them. To the best of our knowledge, neither of these requirements are satisfied in the existing temporal point process literature.\\
In this work we extend the architecture introduced by \citeA{NEURIPS2019_39e4973b} so that it may also deal with earthquake magnitudes. For this we require a model that is dependent on previous event magnitudes, can forecast subsequent magnitudes as well as  forecast earthquakes above a threshold magnitude. Distinguishing between a threshold for the input magnitude and a threshold for the target earthquakes is a problem specific requirement for earthquake forecasting, so does not exist in other works on temporal point processes. We choose to extend \citeA{NEURIPS2019_39e4973b} to give the most flexible representation of the intensity, since they use a fully non-parametric approach compared to other intensity based methods that use a semi-parametric approach. Working with the intensity function rather than directly modelling the likelihood of the next event provides a model that is closer in interpretation to ETAS and provides a natural way to forecast earthquakes above some target threshold magnitude, detailed in section \ref{targeting_events}.\\
Although the use of the ETAS model \cite{ogata1988statistical} has been the dominant way for modelling seismicity in both retrospective and fully prospective forecasting experiments \cite<e.g.>{woessner2011retrospective,taroni2018prospective,cattania2018forecasting,mancini2019improving,mancini2020predictive} as well as in operational earthquake forecasting \cite{marzocchi2014establishment,rhoades2016retrospective,field2017synoptic}, it is restricted to its rigid parametric form. As ETAS only describes the self-exciting nature of seismicity, it cannot capture any kind of inhibition or release of stress such as captured by stress-release models \cite{zheng1991application,xiaogu1994further,bebbington2003linked} or models based on elastostatic stress transfer and Coulomb Rate-and-State (CRS) friction \cite{dieterich1994constitutive}. Furthermore, foreshock activity that differs from ETAS has also been observed \cite{mcguire2005foreshock,brodsky2011spatial,lippiello2012spatial,ogata2014comparing}. Beyond this understanding that ETAS is misspecified, there are also difficulties and inefficiencies with fitting and forecasting. To estimate the intensity, ETAS sums over all previous earthquakes, which requires substantial memory and slows the fitting process and forecasting simulations. For large earthquakes in the past this is important, because their contribution can last more than 100 years \cite{utsu1995centenary}. However, for smaller earthquakes particularly found in enhanced catalogs, one expects the contribution to be close to zero after a far shorter amount of time, making summing over these terms inefficient  \cite{helmstetter2003importance,marsan2008extending}. Furthermore, a particular difficulty with fitting the ETAS model is that there needs to be a reliable estimate of the completeness across the time of the catalog and this needs to be incorporated into the model. Failing to do so will result in biases \cite{hainzl2016rate,zhuang2017data,seif2017estimating}. Methods that attempt to do this either truncate the periods of time where the catalog is most incomplete \cite{kagan1991likelihood,hainzl2008impact}, leading to parameters that can be dominated by a few aftershock sequences, or attempt to model the data incompleteness itself \cite{omi2014estimating,hainzl2016apparent,hainzl2016rate,mizrahi2021embracing}, but this adds additional computational requirements over a standard ETAS model.
\\
To benchmark our proposed neural point process with ETAS, we design forecasting experiments on both synthetic data as well as a new enhanced catalog for the 2016–2017 Amatrice–Visso–Norcia (AVN) seismic sequence. The catalog generated by \citeA{tan2021machine}, containing roughly 900,000 earthquakes, was generated using a deep neural network based phase picker for earthquake arrival times \cite{zhu2019phasenet}. As a result, the size of the catalog increased 10 fold from the routine catalog generated by the Italian National Institute of Geophysics and Volcanology (INGV). The INGV catalog has been used in several retrospective forecasting experiments \cite{ebrahimian2017robust,marzocchi2017earthquake,mancini2019improving, mancini2022use}, but there has yet to be much development of forecasting models using enhanced catalogs such as this one and, given that they contain considerably more earthquakes, investigations into how we can harness these machine learning generated enhanced catalogs are essential. The AVN sequence contains ten $M5 +$ events during a five month period over an 80 km long normal-fault system \cite{mancini2019improving}. The number of large earthquakes as well as the compactness of the region on which they occur make this sequence preferable for testing purely temporal forecasting models which contain no spatial covariates. We do still expect some loss of information by ignoring the spatial covariates, particularly for smaller earthquakes where there is a spatial extent across which earthquakes won't interact.\\
We seek to understand how taking different magnitude thresholds and temporal partitions of our datasets affects the performance of the two models. Through altering these two aspects, we naturally change the amount of data shown to each model so that we may see how sensitive their forecasting performance is to training sample size \cite{wang2010standard}. Through partitioning in time we can see how the performance is affected by the number of major earthquakes that each model is trained on. By altering the magnitude threshold of the input catalog, we seek to improve the predictive skill of forecasts by using the hypothesis that small earthquakes should help to forecast the moderate-to-large earthquakes. Either from a time-independent perspective where large earthquakes are found to nucleate in areas that have a large density of small events \cite{kafka1998well,kafka2000does}, or in time-dependent forecasting (eg. ETAS and CRS) where earthquake triggering is believed to exist at all scales \cite{helmstetter2003importance, marsan2005role, nandan2016systematic}, reducing the threshold of the input catalog generally leads to improved forecasting performance of moderate-to-large events \cite{helmstetter2007high, werner2011high,helmstetter2014adaptive, mancini2022use}.\\
Particularly, \citeA{mancini2022use} consider the same sequence as this study, and compare forecasting results from models trained on several different enhanced catalogs (including the \citeA{tan2021machine} catalog used in this study). They find that \add{the forecasting of M3+ events by }CRS and ETAS models \change{are}{is} not improved by training \remove{them}on the enhanced catalogs\change{,}{.}\add{ When using the same catalog as this study,} \change{however, the new}{they see the} models increase in performance as the input magnitude threshold is lowered from M5 to M3\change{. At}{, but at} the lowest two thresholds M1 and M2, the performance of ETAS is worse than for M3+\remove{when fitting and testing the models with the same catalog used in this study}. Direct comparison of results, however, isn't possible as they report information gains that are for spatio-temporal forecasts using the Poisson assumption of earthquake rates in gridded space-time windows. This study hopes to provide some further insight into the performance of forecasting models using the low magnitude earthquakes found in this catalog and presents neural point processes as a competitive model using such events.
\section{Data}
To benchmark our neural point process against ETAS we conduct forecasting experiments on both real data from the Amatrice-Visso-Norcia sequence as well as synthetic data generated by ETAS. Since we are making comparisons about temporal models, in both catalogs, we remove all spatial covariates.
\subsection{Amatrice-Visso-Norcia High Resolution Catalog}
\begin{figure}
    \centering
    \includegraphics[width=\textwidth]{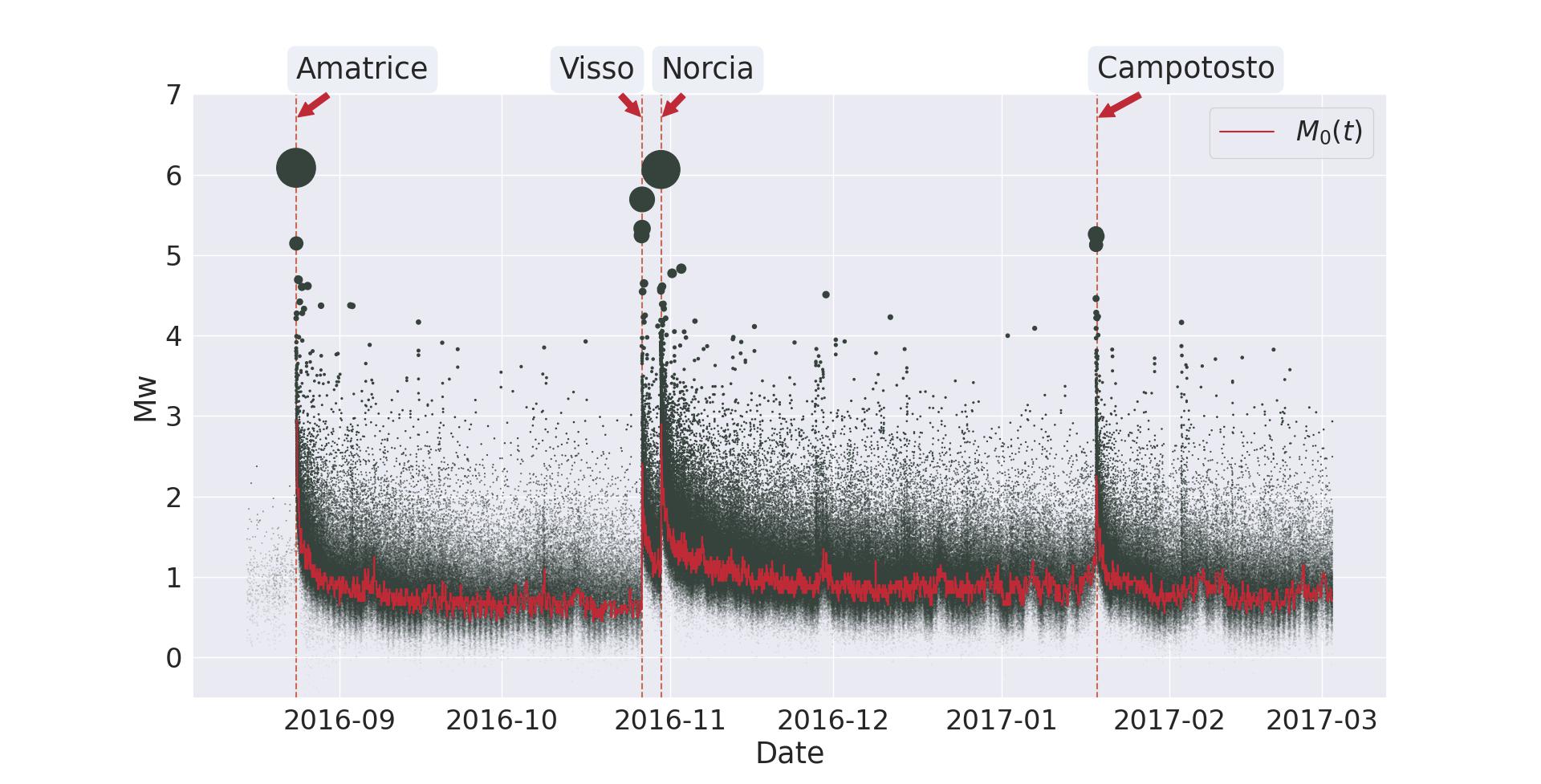}
    \caption{The magnitudes and times of the AVN sequence 2016-2017 \protect\cite{tan2021machine} used to evaluate the performance of the neural and ETAS model. Marked with a dashed red line are the times of the 4 major events of the sequence. The size of the points are plotted on a log scale corresponding to $M_\text{w}$. An estimate of the temporal completeness $M_0(t)$ is plotted using the maximum curvature method \protect\cite{wiemer2000minimum}.}
    \label{fig:the_dataset}
\end{figure}

On the 24th of August 2016 a $\text{M}_\text{w}6.0$ earthquake was recorded near the town of Amatrice in northern Lazio, central Italy. It was followed by a $\text{M}_\text{w}$5.9 near the town of Visso on the 26th of October and a $\text{M}_\text{w}$6.5 near the town of Norcia four days later. Finally, in January 2017, four events between $\text{M}_\text{w}$5.0 and $\text{M}_\text{w}$5.5  struck the Campotosto area. Figure \ref{fig:the_dataset} depicts the evolution of this seismic sequence over time.\\
The INGV produced a routine catalog for the 1 year period containing this sequence \cite{chiaraluce20172016}. Their catalog contains roughly 82,000 events with a completeness of $Mc = 2.3$ \cite{mancini2019improving}. With the use of a neural network based phase picker to determine P and S-wave arrival times, an enhanced catalog has been created for the same earthquake sequence \cite{tan2021machine}. This catalog contains around 900,000 events and has an overall value of completeness of the catalog $Mc = 0.2$. We estimated the time varying completeness of this catalog using the maximum curvature method \cite{wiemer2000minimum} with samples of 300 events and can see clear variation in completeness particularly following large magnitude earthquakes. This approximate method for the completeness is only used to show the variability across the catalog and is not used directly in any modelling.

\subsection{Synthetic Catalog}
For the forecasting experiment of synthetic ETAS data, we generate a dataset using the simulator by \citeA{mizrahi2021effect}, with uniform background intensity $\mu$ and triggering function,
\begin{linenomath*}
\begin{equation*}
    g(t,x,y,m) = \frac{k_0 e^{a(m-M_0)}}{\frac{(t+c)^{1+\omega}}{e^{t/\tau}}((x^2+y^2) + d e^{\gamma(m-M_0)})^{1+\rho}}.
\end{equation*}
\end{linenomath*}
The ETAS parameters ($\log_{10} \mu =  -6.6,\ \log_{10} k_0 = -3.15,\ a = 2.85,\ \log_{10} c = -2.95,\ \omega = -0.03,\ \log_{10} \tau = 3.99,\ \log_{10} d =-0.35,\ \gamma = 1.22,\ \rho = 0.51,\ M_0 = 1\ M_\text{w}$) are taken close to \citeA{mizrahi2021effect} with higher background rate to account for the lower $M_0$. The resulting data\remove{t}set of roughly 250,000 events comes from removing all the spatial covariates.\\
We also generate a second synthetic dataset from the first by emulating short-term aftershock incompleteness using the time-dependent formula from \citeA{helmstetter2006comparison},
\begin{linenomath*}
\begin{equation*}
    M_0(M,t) = M - 4.5 - 0.75 \log(t),
\end{equation*}
\end{linenomath*}
where $M$ is the mainshock magnitude. Events below the function are removed using the six largest events as mainshocks in this synthetic catalog.
\section{Theoretical Background}
In this section we briefly introduce the theory of neural point processes. We begin with the basic theory of temporal point processes and show how to use this to construct neural point processes for temporal forecasts only. In Section 4, we extend a temporal neural point process to a marked neural process. 
\subsection{Point Processes}
A temporal point process \cite{daley2003introduction} is a stochastic process that generates a sequence of discrete events at times $\{t_i\}_{i=1}^n$ in a given observation interval $[0,T]$. The process is characterised by its conditional intensity function $\lambda(t|H_t)$, which gives the expected number of events in a small interval about $t$ conditioned on the event history $H_t = \{t_i|t_i<t\}$:
\begin{linenomath*}
\begin{equation*}
    \lambda(t|H_t)dt = \mathbb{E}\left[N([t+dt])|H_t\right] . 
\end{equation*}
\end{linenomath*}
The intensity function \cite{rasmussen2018lecture} completely defines the point process and can take a variety of functional forms. The most basic form is the stationary Poisson process \cite{daley2003basic} which assumes that all events are independent of each other, and the conditional intensity function is constant. Self-exciting point processes assume that events increase the likelihood of subsequent events, with a popular class of these processes being the Hawkes process \cite{hawkes1971spectra}. The Hawkes process is defined by its conditional intensity $\lambda(t|H_t) = \mu + \sum_{t_i<t}g(t-t_i)$, where $g(s)$ is a kernel function defining how past events trigger subsequent events.\\
With the conditional intensity function specified we can write an expression for the log-likelihood of observing a sequence of events $\{t_i\}_{i=1}^n$,
\begin{linenomath*}
\begin{equation}\label{eq:unmarked_likelihood}
    \log L(\{t_i\}) = \sum_{i=1}^n\left[ \log \lambda(t_i|H_{t_i})\ -\int_{t_{i-1}}^{t_i} \lambda(t|H_t)dt    \right].
\end{equation}
\end{linenomath*}
We assume that the observation interval is $[t_1,t_n]$ to make the algebra in the remainder more compact. It is trivial to relax this to an arbitrary interval $[0,T]$.\\
Temporal point processes can be extended to incorporate marks. A marked point process is stochastic process that generates events paired with a mark, $\{t_i,m_i\}_{i=1}^n \in \left(\mathbb{R}_{>0} \times \mathcal{M}\right)$. In our setting this represents the occurrence times of earthquakes with their corresponding magnitudes. A marked point process is defined by its conditional intensity function,
\begin{linenomath*}
\begin{equation*}
    \lambda(t,m|H_t)dtdm = \mathbb{E}\left[N(dt \times dm)|H_t\right],
\end{equation*}
\end{linenomath*}
with the log-likelihood of observing a marked sequence of events given by
\begin{linenomath*}
\begin{equation}\label{eq:marked_likelihood}
\log L(\{t_i,m_i\}) = \sum_i \left[ \log \lambda(t_i,m_i|H_t)\ -\int_{t_{i-1}}^{t_i} \int_{\mathcal{M}} \lambda(t,m|H_t)dmdt    \right].
\end{equation}
\end{linenomath*}
\subsection{Neural Point Processes}
The most common form of neural point processes seeks to obtain a compact representation of the event history through the use of an RNN \cite{du2016recurrent}. In this approach, an input representing the inter-event times $\tau_i = t_i-t_{i-1}$ is first fed into the RNN. A hidden state $\textbf{h}_i$ of the RNN is updated
\change{\protect\begin{linenomath*}
\begin{equation*}
    \textbf{h}_i = f(W^h\textbf{h}_{i-1} + \textbf{w}^\tau \tau_i + \textbf{b}^h)
\end{equation*}
\end{linenomath*}}{\protect\begin{linenomath*}
\begin{equation*}
    \textbf{h}_i = \sigma(W^h\textbf{h}_{i-1} + \textbf{w}^\tau \tau_i + \textbf{b}^h)
\end{equation*}
\end{linenomath*}}
where $\{W^h,\textbf{w}^\tau, \textbf{b}^h\}$ are learnable parameters, and \change{$f$}{$\sigma$} is an activation function. The conditional intensity function is then formulated as a function of the elapsed time from the most recent event and \add{is dependent on }the hidden state of the RNN,
\change{\protect\begin{linenomath*}
\begin{equation*}
    \lambda(t|H_t) = \phi(t-t_i,\textbf{h}_i),
\end{equation*}
\end{linenomath*}}{\protect\begin{linenomath*}
\begin{equation*}
    \lambda(t|H_t) = \phi(t-t_i|\textbf{h}_i),
\end{equation*}
\end{linenomath*}}
where $\phi$ is a non-negative function referred to as the hazard function\add{ and $t_i$ is the time of the most recent event}. To avoid having to numerically integrate the intensity function directly, \citeA{NEURIPS2019_39e4973b} model the integral of the hazard function with a fully connected neural network
\change{\protect
\begin{linenomath*}
\begin{equation*}
    \Phi(\tau_i,\textbf{h}_i) = \int_0^{\tau_i} \phi(s,\textbf{h}_i)ds.
\end{equation*}
\end{linenomath*}}{
\protect \begin{linenomath*}
\begin{equation*}
    \Phi(\tau|\textbf{h}_i) = \int_0^{\tau} \phi(s|\textbf{h}_i)ds.
\end{equation*}
\end{linenomath*}
}

With the construction of the model in this way, the log-likelihood of observing a sequence of event times (\ref{eq:unmarked_likelihood}) can be written as:
\change{\protect
\begin{linenomath*}
\begin{equation*}
\log L(\{t_i\}) = \sum_i \left[ \log \frac{\partial}{\partial \tau_i}\Phi(\tau_i,\textbf{h}_i ) - \Phi(\tau_i,\textbf{h}_i)\right].
\end{equation*}
\end{linenomath*}}{\protect
\begin{linenomath*}
\begin{equation*}
\log L(\{t_i\}) = \sum_i \left[ \log \frac{\partial}{\partial \tau}\Phi(\tau_i|\textbf{h}_i ) - \Phi(\tau_i|\textbf{h}_i)\right].
\end{equation*}
\end{linenomath*}
}
\section{Methods}
Since the neural point process introduced by \citeA{NEURIPS2019_39e4973b} is purely temporal, to model seismicity we must extend their model to our requirements. Particularly we require that forecasts be dependent on the history of both times and magnitudes, as magnitudes are an important predictor of seismicity \cite{utsu1970aftershocks,utsu1971aftershocks}. We also require a forecast over the next magnitude where, unlike the Gutenberg-Richter law \cite{gutenberg1936magnitude} routinely assumed in the ETAS framework, this is also dependent on the history of events. Finally, we require that we may make forecasts of earthquakes above some target magnitude threshold despite a dependence on earthquakes below that target threshold. In Section \ref{Continuously Marked Neural Point Process}, we first extend the structure of the neural network by \citeA{NEURIPS2019_39e4973b} to maximise the likelihood of observing a marked sequence of events, including constructing a time-history dependent magnitude distribution. In Section \ref{targeting_events}, we show how we adjust this new structure to target events above a magnitude threshold. \\
To aid in the development of more flexible forecasting models, we will make both the dataset and models used in this study available after publication on \url{https://github.com/ss15859/Neural-Point-Process}.
\subsection{Continuously Marked Neural Point Process}\label{Continuously Marked Neural Point Process}
\begin{figure}
    \centering
    \includegraphics[width=\textwidth]{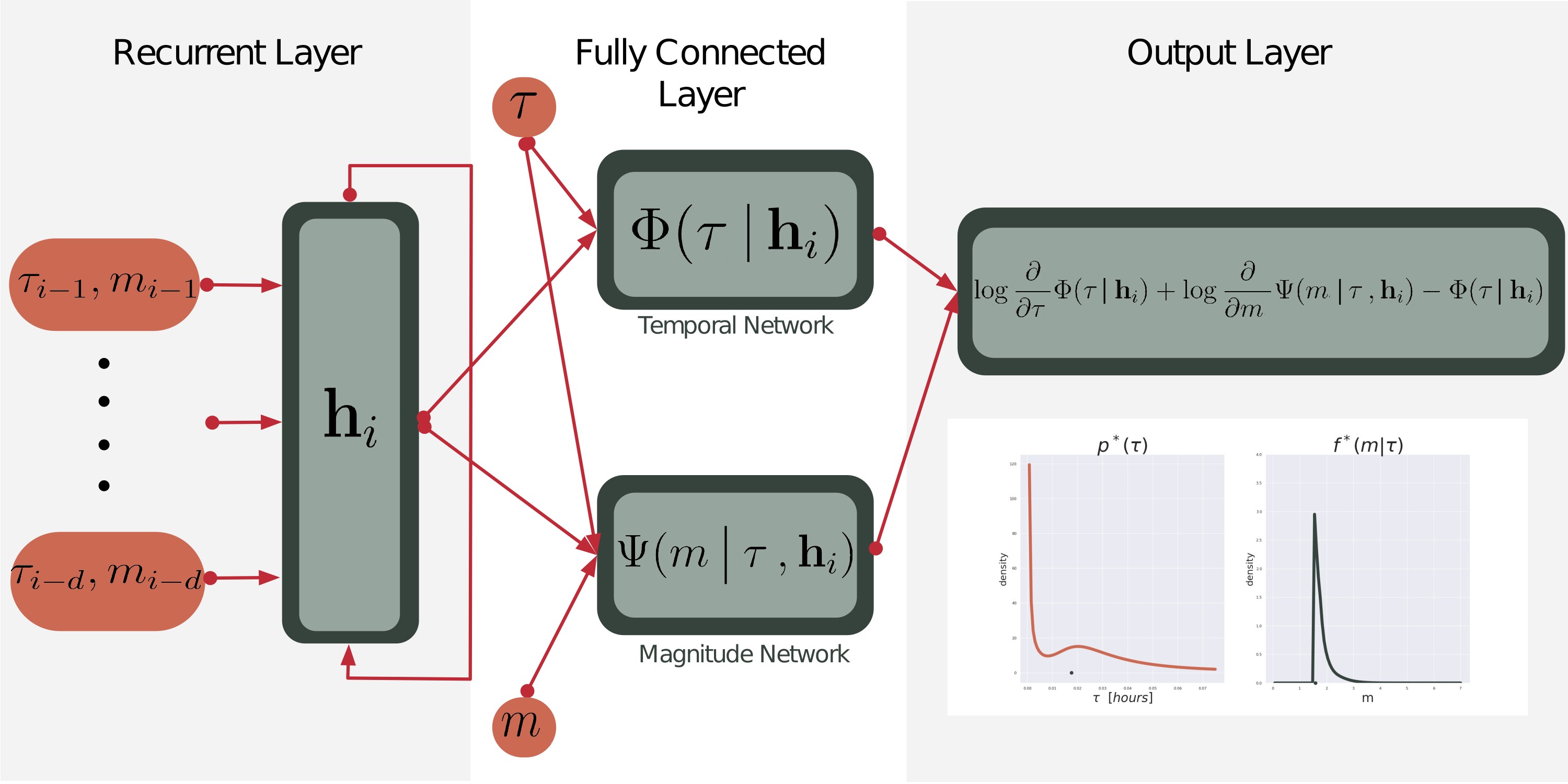}
    \caption{The proposed network comprises four sections. First, the inter-event times and magnitudes of the last d events are fed into a recurrent section consisting of 64 recurrent units. The output of this section is fed into two fully connected sections where it is combined with the next inter-event time \add{$\tau$ }for the temporal network and additionally with the next magnitude \add{$m$ }for the magnitude network. The outputs of both these sections are combined to formulate the log-likelihood of the next inter-event time and magnitude\add{ $\{\tau,m\}$}. We can separate the temporal and magnitude terms in this likelihood to give point evaluations of the density of the next inter-event time and conditional density of the next magnitude. The dependence structure of the point process is expressed by the connections between sections in the network.}
    \label{fig:network_diag}
\end{figure}

We begin with the factorisation of the joint conditional intensity function into its marginal intensity and conditional density function, following \citeA{daley2003introduction},
\begin{linenomath*}
\begin{equation*}
    \lambda^*(t,m) = \lambda^*(t) f^*(m|t),
\end{equation*}
\end{linenomath*}
where $\lambda^*(t)$ is the marginal conditional intensity function of $t$, and $f^*(m|t)$ is the conditional density function of the mark at time $t$. Both of these functions are dependent on the history $H_t$, here denoted by the asterisk *. To construct the likelihood for the marked sequence we model these two functions separately.\\
With the factorisation, the expression for the log-likelihood of observing the marked sequence of events (\ref{eq:marked_likelihood}) becomes,
\begin{linenomath*}
\begin{equation}\label{eq:marked_likelihood_2}
    \log L(\{t_i,m_i\}) = \sum_i \left[ \log \lambda^*(t_i) + \log f^*(m_i|t_i) - \int_{t_{i-1}}^{t_i}  \lambda^*(t) dt\right].
\end{equation}
\end{linenomath*}
Now with a two dimensional input, the hidden state of the RNN is updated as a linear combination of the inter-event times and magnitudes. This is the continuous mark extension to the RNN update from \citeA{du2016recurrent},
\change{\protect\begin{linenomath*}
\begin{equation*}
    \textbf{h}_i = f(W^h\textbf{h}_{i-1} + \textbf{w}^\tau \tau_{i-1}+ \textbf{w}^m m_{i-1} + \textbf{b}^h),
\end{equation*}
\end{linenomath*}}{\protect\begin{linenomath*}
\begin{equation*}
    \textbf{h}_i = \sigma(W^h\textbf{h}_{i-1} + \textbf{w}^\tau \tau_{i-1}+ \textbf{w}^m m_{i-1} + \textbf{b}^h),
\end{equation*}
\end{linenomath*}}
where $\textbf{w}^m$ is an additional learnable parameter.\\
The marginal intensity function is formulated as a function of the elapsed time from the most recent event and \add{is dependent on }the hidden state of the RNN \cite{du2016recurrent},\change{\protect
\begin{linenomath*}
\begin{equation*}\label{eq:NN_modeling_choice}
    \lambda(t_i|H_{t_i}) = \phi(\tau_i=t_i-t_{i-1},\textbf{h}_{i}),
\end{equation*}
\end{linenomath*}}{\protect
\begin{linenomath*}
\begin{equation}\label{eq:NN_modeling_choice}
    \lambda(t|H_{t}) = \phi(\tau=t-t_{i}|\textbf{h}_{i}),
\end{equation}
\end{linenomath*}
}
where $\phi$ is a non-negative function referred to as the hazard function\add{ and $t_i$ is the time of the most recent event}. Following \citeA{NEURIPS2019_39e4973b}, we model the cumulative hazard function using a fully connected neural network,
\change{\protect
\begin{linenomath*}
\begin{equation*}
    \Phi(\tau_i,\textbf{h}_i) = \int_0^{\tau_i} \phi(s\textbf{h}_i)ds.
\end{equation*}
\end{linenomath*}}{
\protect \begin{linenomath*}
\begin{equation*}
    \Phi(\tau|\textbf{h}_i) = \int_0^{\tau} \phi(s|\textbf{h}_i)ds.
\end{equation*}
\end{linenomath*}
}
which allows us to differentiate this with respect to \change{$\tau_i$}{$\tau$} to extract the hazard function. The derivative is easily obtained through automatic differentiation \cite{van2018automatic}, which is available in all neural network packages.\\
We now also formulate the conditional density function of the mark at time $t$ as a function of the current mark\change{,}{. This is dependent on} the time since the most recent event and the hidden state of the RNN,
\change{\protect\begin{linenomath*}
\begin{equation*}
    f(m_i|t_i,H_{t_i}) = \psi(m_i,\tau_i,\textbf{h}_i).
\end{equation*}
\end{linenomath*}}{\protect
\begin{linenomath*}
\begin{equation*}
    f(m|t,H_{t_i}) = \psi(m|\tau,\textbf{h}_i).
\end{equation*}
\end{linenomath*}
}
We again model its cumulative distribution with a fully connected neural network,
\change{\protect\begin{linenomath*}
\begin{equation*}
    \Psi(m_i,\tau_i,\textbf{h}_i) = \int_0^{m_i} \psi(\mu,\tau_i,\textbf{h}_i) d\mu.
\end{equation*}
\end{linenomath*}}{\protect
\begin{linenomath*}
\begin{equation*}
    \Psi(m|\tau,\textbf{h}_i) = \int_0^{m} \psi(\mu|\tau,\textbf{h}_i) d\mu.
\end{equation*}
\end{linenomath*}
}
Although this integral does not feature in the expression for the log-likelihood, we still opt for this approach over directly modelling the density function with a neural network. This follows from work on neural density estimation where positive weights can be enforced in the network to capture the positivity and monotonicity of cumulative distribution functions \cite{chilinski2020neural}. We can then obtain the density function again through automatic differentiation.\\
We can now write the log-likelihood as:
\change{\protect
\begin{align*}
    \log L(\{t_i,m_i\}) &= \sum_i \left[ \log \lambda^*(t_i) + \log f^*(m_i|t_i) - \int_{t_{i-1}}^{t_i}  \lambda^*(t) dt\right]\\
    & = \sum_i \left[  \log \phi(\tau_i,\textbf{h}_i)   +  \log \psi(m_i,\tau_i,\textbf{h}_i)- \int_0^{t_{i}-t_{i-1}}  \phi(t,\textbf{h}_i) dt\right]\\
    &= \sum_i \left[ \log \frac{\partial}{\partial \tau_i}\Phi(\tau_i,\textbf{h}_i )     +  \log \frac{\partial}{\partial m_i}\Psi(m_i,\tau_i,\textbf{h}_i)- \Phi(\tau_i,\textbf{h}_i)\right].
\end{align*}
}{\protect\begin{linenomath*}
\begin{align}
    \log L(\{t_i,m_i\}) &= \sum_i \left[ \log \lambda^*(t_i) + \log f^*(m_i|t_i) - \int_{t_{i-1}}^{t_i}  \lambda^*(t) dt\right]\\
    & = \sum_i \left[  \log \phi(\tau_i|\textbf{h}_i)   +  \log \psi(m_i|\tau_i,\textbf{h}_i)- \int_0^{t_{i}-t_{i-1}}  \phi(t|\textbf{h}_i) dt\right]\\
    &= \sum_i \left[ \log \frac{\partial}{\partial \tau}\Phi(\tau_i|\textbf{h}_i )     +  \log \frac{\partial}{\partial m}\Psi(m_i|\tau_i,\textbf{h}_i)- \Phi(\tau_i|\textbf{h}_i)\right].\label{eq:NN_log_lik}
\end{align}
\end{linenomath*}}
We model both the cumulative hazard function $\Phi$, and the conditional distribution function of the marks $\Psi$, using a feed-forward neural network. The network depicted in Figure \ref{fig:network_diag} consists of four component parts. The first part is the recurrent section, which finds an encoding of the history of the point process. The output of the recurrent section \add{$\textbf{h}_i$ }passes into two fully connected components. One models the integral of the intensity function, \change{taking also as input}{this is a function of} the time of the next event\add{ $\tau$}. The other component models the integral of the conditional density function of the next mark, \add{this is }dependent on the next time\add{ $\tau$}, \change{taking both of these values as input}{but is a function of the next mark $m$}. \remove{The structure of the network describes the dependence relationship of the point process, represented by a connection between sections in the diagram. For example, the magnitude network models the conditional cumulative magnitude distribution: a function that depends on the observed magnitude $m_i$, time since the last event $\tau_i$ and the history $\textbf{h}_i$.}\\
Since passing long sequences into recurrent neural networks can often lead to exploding or vanishing gradient problems \cite{hochreiter1998vanishing}, we do not pass the whole history of the point process into the recurrent section, but the past $d$ events. This implies that we use only the past $d$ events to forecast the next event. Thus, this model is learning to estimate the intensity given a recent history of $d$ events. This hyperparameter is kept the same as \citeA{NEURIPS2019_39e4973b} at $d=20$. A naive tuning search found no significant improvement at larger values of $\{50,100,200,500\}$. This difference to the full-history ETAS model is discussed in Section \ref{Aprox_ETAS}. \\
We enforce positive weights in both fully connected sections of the network to capture the positivity and monotonicity that is required by both cumulative functions.
In the final component of the network we formulate the output as the log-likelihood of observing the pair \change{$\{\tau_i,m_i\}$}{$\{\tau,m\}$}. To construct this output, one backward pass is performed to calculate the derivatives with respect to the next time and the next magnitude found in eq (\ref{eq:NN_log_lik}). This output is exactly what is maximised to learn the weight parameters, during a second backwards pass. \\
\subsection{Target Events}\label{targeting_events}
The growth in machine learning generated catalogs from their predecessors is found through detecting events of magnitude much lower than previously possible. However, operationally, we may not care about the forecasting of these smaller events if it is the larger ones that are the most hazardous. The aim is therefore to use the smaller events to forecast earthquakes above some target threshold. In this section we outline how this is done for both ETAS and the neural model. We hereafter call events above the target magnitude threshold $M_d$ target events. \\
Let $\{(t_i,m_i)\}_{i=1}^n \in [0,T] \times \mathcal{M}$ be the entire sequence of events, complete down to $M_0$. We seek to make forecasts of events above magnitude $M_d$. This corresponds to the sequence:
\begin{linenomath*}
\begin{equation*}
   \{ (t_j,m_j) : m_j \geq M_d \}_{j=1}^k.
\end{equation*}
\end{linenomath*}
To ease in the distinction between `all events' and the target events, we subscript the former with $i$ and the latter with $j$.\\
Let $\lambda_0(t,m|H_t)$ denote the joint intensity function of all events above $M_0$. We seek to learn the intensity of events above $M_d$, denoted $\lambda_d(t,m|H_t)$, where the history $H_t$ contains all events $\{(t_i,m_i)\}_{i: t_i <t}$. The ground intensity above the target threshold is found by marginalising the joint intensity over the target magnitude region,
\begin{linenomath*}
\begin{equation*}
   \lambda_d(t|H_t) = \int_{M_d}^\infty \lambda_0(t,m|H_t)dm.  
\end{equation*}
\end{linenomath*}
The log likelihood of target events is then given by:
\begin{linenomath*}
\begin{equation*}
    \log L(\{t_j,m_j\}) = \sum_{j: m_j\geq M_d} \left( \log \lambda_d(t_j,m_j|H_{t_j})    - \int^{t_j}_{t_{j-1}}  \lambda_d(t|H_t) dt \right).
\end{equation*}
\end{linenomath*}
For ETAS, the rate above magnitude $M_d$ is a fraction of the rate above $M_0$, due to the independent distribution for magnitudes,
\change{\protect\begin{linenomath*}
\begin{equation*}
\lambda_d(t|H_t) = \int_{M_d}^\infty \lambda_0(t,m)dm  =\left( \int_{M_d}^\infty f(m) dm \right) \lambda_0 (t|H_t) = p_d \cdot \lambda_0 (t|H_t),
\end{equation*}
\end{linenomath*}}{\protect\begin{linenomath*}
\begin{equation*}
\lambda_d(t|H_t) = \int_{M_d}^\infty \lambda_0(t,m)dm  =\left( \int_{M_d}^\infty f_{GR}(m) dm \right) \lambda_0 (t|H_t) = p_d \cdot \lambda_0 (t|H_t),
\end{equation*}
\end{linenomath*}}
where \change{$f(m)$}{$f_{GR}(m)$} is the Gutenburg-Richter law and $p_d$ is simply the probability that $m \geq M_d$. Therefore the expression for the likelihood is relatively simple,
\change{\protect\begin{linenomath*}
\begin{equation*}
    \log L(\{t_j,m_j\}) = \sum_{j: m_j \geq M_d} \left( \log p_d \cdot  \lambda_0(t_j,m_j|H_{t_j})    - \int^{t_j}_{t_{j-1}} p_d \cdot \lambda_0(t|H_t)dt \right).
\end{equation*}
\end{linenomath*}}{\protect\begin{linenomath*}
\begin{equation*}
    \log L(\{t_j,m_j\}) = \sum_{j: m_j \geq M_d} \left[ \log \left(p_d \cdot  \lambda_0(t_j,m_j|H_{t_j})\right)    - \int^{t_j}_{t_{j-1}} p_d \cdot \lambda_0(t|H_t)dt \right].
\end{equation*}
\end{linenomath*}}
For the neural model, we make use of the fact that the integral of the intensity function between target events\add{, $\{ (t_j,m_j) : m_j \geq M_d \}_{j=1}^k$, } can be expressed as a sum of disjoint integrals between all events\add{  $\{(t_i,m_i)\}_{i=1}^n$},
\change{\protect\begin{linenomath*}
\begin{align*}
\log L(\{t_j,m_j\}) &= \sum_{j: m_j \geq M_d} \left[ \log \lambda_d(t_j,m_j|H_{t_j}) - \int^{t_j}_{t_{j-1}}  \lambda_d(t|H_t) dt \right] \\
&= \sum_{j: m_j \geq M_d} \left[ \log \lambda_d(t_j|H_{t_j}) + \log f_d(m_j|t_j,H_{t_j})\right]  - \int^{t_k}_{t_{0}}  \lambda_d(t|H_t) dt  \\
&= \sum\limits_{\substack{i:m_i \geq M_0 \\ t_i \leq t_k}} \left[\left( \log \lambda_d(t_i|H_{t_i}) + \log f_d(m_i|t_i,H_{t_i}) \right)\mathbb{\mathbf{I}}\{m_i \geq M_d\}    - \int^{t_i}_{t_{i-1}}  \lambda_d(t|H_t) dt \right] \\
    &= \sum\limits_{\substack{i:m_i \geq M_0 \\  t_i \leq t_k}}\left[ \left(\log \frac{\partial}{\partial \tau_i}\Phi(\tau_i,\textbf{h}_i )     +  \log \frac{\partial}{\partial m_i}\Psi(m_i,\tau_i,\textbf{h}_i)\right)\mathbb{\mathbf{I}}\{m_i \geq M_d\} -  \Phi(\tau_i,\textbf{h}_i) \right], 
\end{align*}
\end{linenomath*}}{\protect\begin{linenomath*}
\begin{align}
\log L(\{t_j,m_j\}) &= \sum_{j: m_j \geq M_d} \left[ \log \lambda_d(t_j,m_j|H_{t_j}) - \int^{t_j}_{t_{j-1}}  \lambda_d(t|H_t) dt \right] \label{eq:OG}\\
&= \sum_{j: m_j \geq M_d} \left[ \log \lambda_d(t_j|H_{t_j}) + \log f_d(m_j|t_j,H_{t_j})\right]  - \int^{t_k}_{t_{0}}  \lambda_d(t|H_t) dt \label{eq:js} \\
&= \sum\limits_{\substack{i:m_i \geq M_0 \\ t_i \leq t_k}} \left[\left( \log \lambda_d(t_i|H_{t_i}) + \log f_d(m_i|t_i,H_{t_i}) \right)\mathbb{\mathbf{I}}\{m_i \geq M_d\}    - \int^{t_i}_{t_{i-1}}  \lambda_d(t|H_t) dt \right] \label{eq:is}\\
    &= \sum\limits_{\substack{i:m_i \geq M_0 \\  t_i \leq t_k}}\left[ \left(\log \frac{\partial}{\partial \tau}\Phi(\tau_i|\textbf{h}_i )     +  \log \frac{\partial}{\partial m}\Psi(m_i|\tau_i,\textbf{h}_i)\right)\mathbb{\mathbf{I}}\{m_i \geq M_d\} -  \Phi(\tau_i|\textbf{h}_i) \right], \label{eq:objective_func}
\end{align}
\end{linenomath*}}
where between (\ref{eq:OG}) and (\ref{eq:js}) we have factorised the joint intensity of events above $M_d$, $\lambda_d(t,m|H_{t})$, into the ground intensity above the target threshold, $\lambda_d(t|H_t)$, and the distribution of the next magnitude above the target threshold given the time and the history, $f_d(m|t,H_{t})$.
Between (\ref{eq:js}) and (\ref{eq:is}) we changed the summation from being over target events to being over all events\change{ .}{ by adding the indicator function \protect$\mathbf{I}\{m_i \geq M_d\}$. The integral in \protect (\ref{eq:js}) becomes the sum of integrals in (\protect\ref{eq:is}) through a decomposition into disjoint integrals between all events  $\{(t_i,m_i)\}_{i=1}^n$.} Thus the neural model may target events by adding the indicator function $\mathbb{\mathbf{I}}\{m_i \geq M_d\}$ to the expression for the log-likelihood. Now the hazard function models the rate above $M_d$ as a function of the time from the last event (of any magnitude),
\change{\protect\begin{linenomath*}
\begin{equation*}
    \lambda_d(t_i|H_{t_i}) = \phi(\tau_i=t_i-t_{i-1},\textbf{h}_{i}).
\end{equation*}
\end{linenomath*}}{\protect\begin{linenomath*}
\begin{equation*}
    \lambda_d(t|H_{t}) = \phi(\tau=t-t_{i}|\textbf{h}_{i}).
\end{equation*}
\end{linenomath*}}
\subsection{Experimental Design}
\begin{figure}
    \centering
    \includegraphics[width=0.9\textwidth]{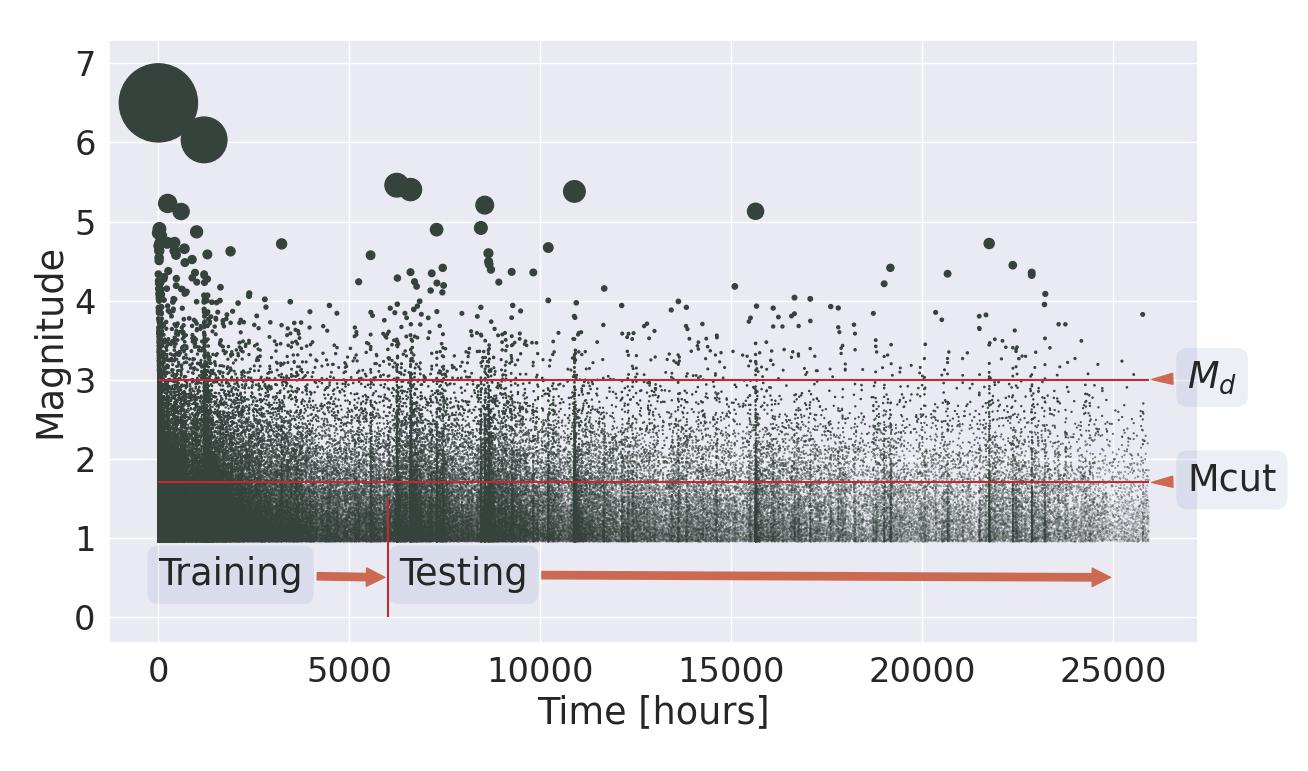}
    \caption{The synthetic catalog with an outline of the training and testing procedure. We train up to a fixed point in time in the catalog, following which the remainder of the catalog is used for testing. We vary the value of the threshold for the input catalog (Mcut) and keep fixed the value of the target threshold ($M_d$).}
    \label{fig:experimental_procedure}
\end{figure}

For both the synthetic data and real data we apply the same training and testing procedure illustrated in Figure \ref{fig:experimental_procedure}. At a fixed point in time along the sequence we set a marker and train on data up to that point. Following that, the remainder of the sequence will be used as the test set. For the synthetic catalogs this is done at one single point in time, whereas for the Central Apennines sequence we make three partitions - each just before the Visso, Norcia and Campotosto earthquakes. By making these partitions we can see how the performance of each model is affected by the number of training datapoints as well as the number of  major earthquakes.\\

We seek to understand how different magnitude thresholding affects the performance of each of the models. For each of the partitions, we look at the performance of both models as the magnitude threshold of the input catalog is lowered, a parameter we refer to as Mcut. For the AVN catalog and incomplete synthetic catalog, this includes lowering Mcut into periods where Mcut$< M_0(t)$. We keep fixed the magnitude of events we wish to target at $\text{M}_d = 3$ Mw.\\
Both models are trained by maximum likelihood estimation (MLE) on the training dataset. For ETAS we use the intensity function defined by \citeA{ogata1988statistical} and maximise the likelihood through Nelder-Mead optimisation, chosen for its robustness. For the neural model, we maximise the likelihood defined in equation (\ref{eq:objective_func}) through ADAM optimisation \cite{kingma2014adam} written in Tensorflow \cite{tensorflow2015-whitepaper}.\\
We compare the two models' performance through the log-likelihood of the events in the testing set. We separate the temporal and magnitude terms in the likelihood equation (\ref{eq:marked_likelihood_2}), to analyse their predictive skills on the two target variables separately. To compare the performance across different magnitude thresholds, we also fit a homogeneous Poisson model. For the temporal log-likelihood we can therefore present the log-likelihood gain from a benchmark Poisson model, whereas for the magnitude forecasts, simply the log-likelihood is reported. We shall refer to both performance metrics as log-likelihood scores and to make general comparisons across the models, we compare the mean log-likelihood score per earthquake as well as construct a $95\% $ confidence interval to assess the variability. The confidence interval is constructed \change{through a zero mean normal assumption on the distribution of the log-likelihood scores with unknown variance. This is the same formulation as the T-test from rhoades2011efficient}{with 1000 bootstrap samples of the log-likelihood scores}.
\section{Results}
\subsection{Synthetic Data}
\begin{figure}
    \centering    
    \includegraphics[width=0.8\textwidth]{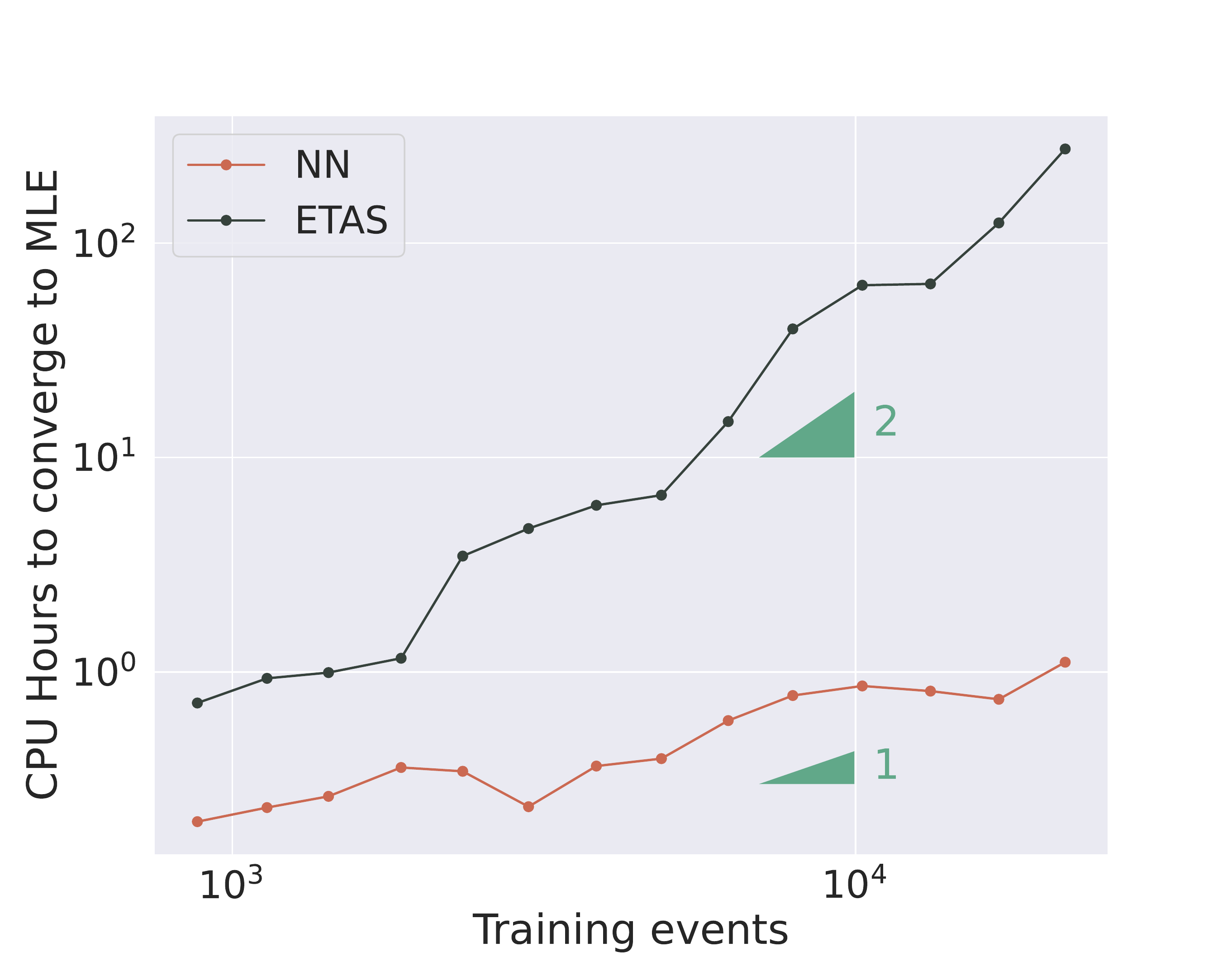}
    \caption{Time on a single CPU required to train each of the models as the training size increases. Each model is trained by maximising the likelihood of the training data.}
    \label{fig:train_time}
\end{figure}

\begin{figure}
    \centering
    \includegraphics[width=\textwidth]{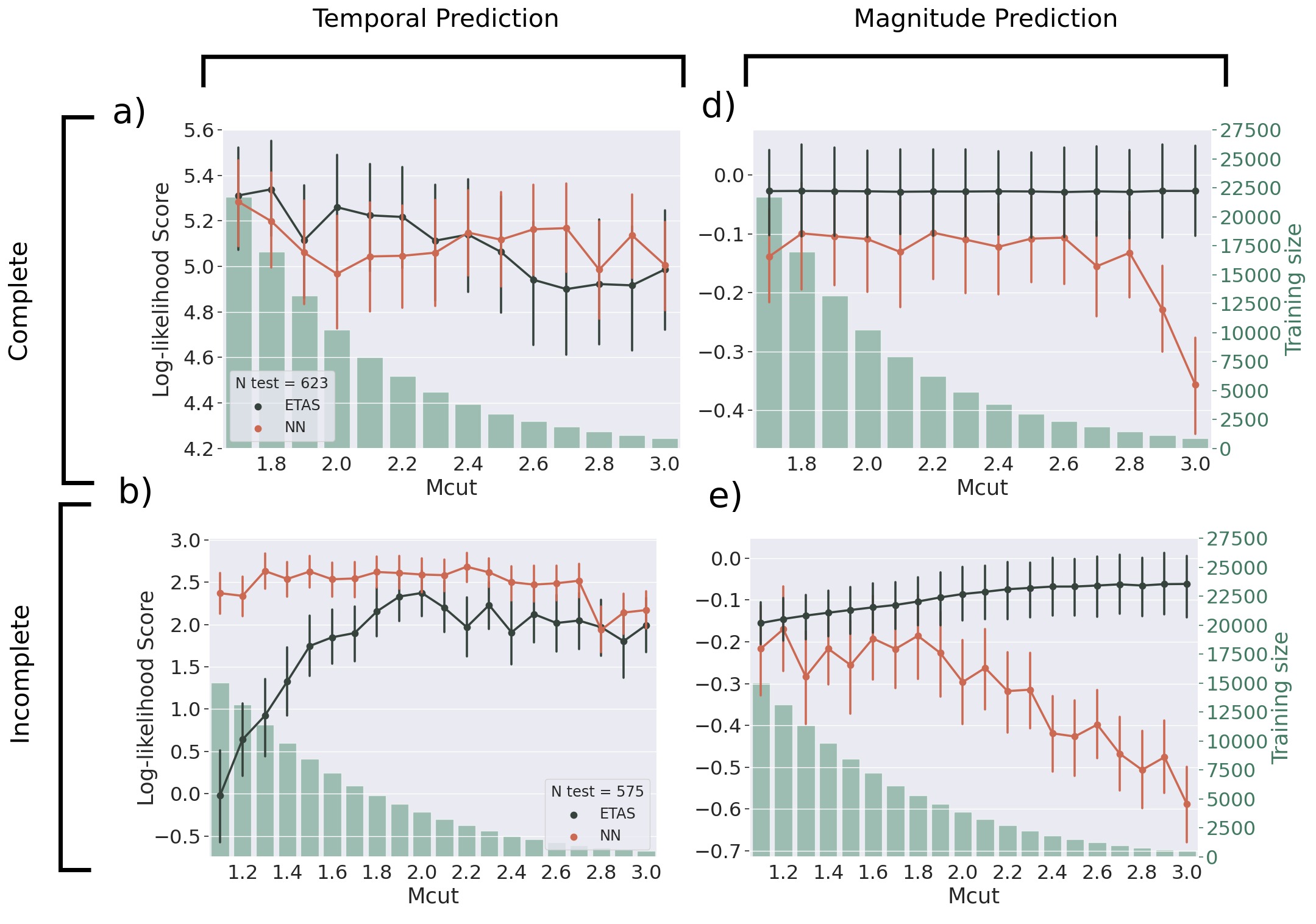}
    \caption{Results from the synthetic tests. 95 \% confidence intervals for the log-likelihood scores for each model as a function of Mcut (the magnitude threshold of the input catalog). The size of the training set is displayed in the green barplot; the size of the testing set in the legend. a) temporal log-likelihood gain from Poisson for the complete synthetic catalog. b) temporal log-likelihood gain for the incomplete catalog. c) magnitude log-likelihood for the complete catalog. d) magnitude log-likelihood for the incomplete catalog.}
    \label{fig:simulated_log_scores}
\end{figure}

\begin{figure}
    \centering
    \includegraphics[width=0.7\textwidth]{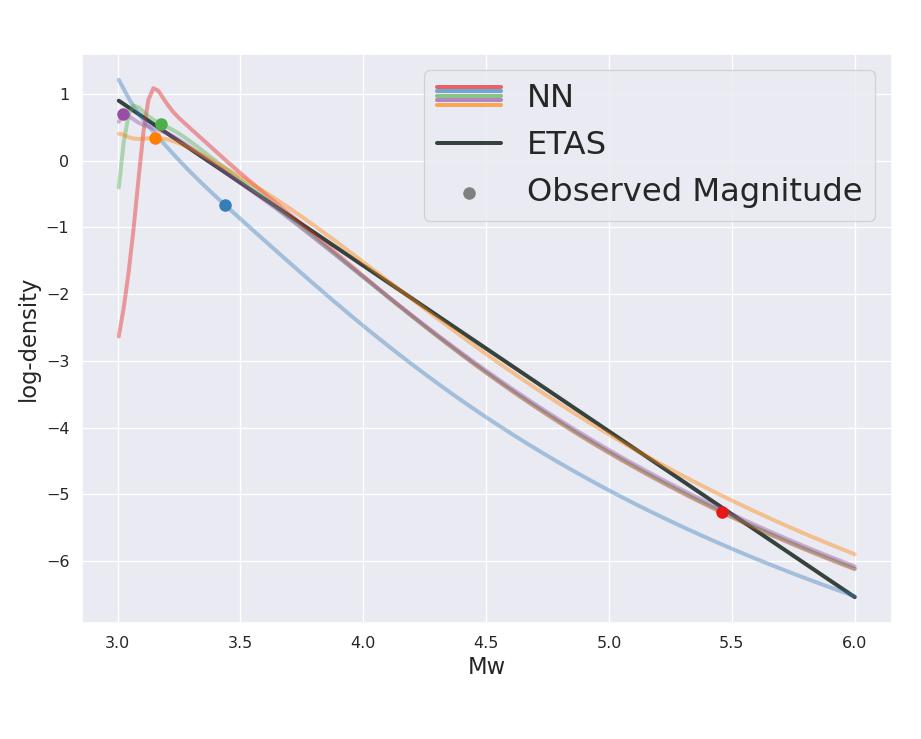}
    \caption{\add{Five examples of the forecasted magnitude distributions from the complete synthetic catalog tests at Mcut = 1.7 compared with the ETAS Gutenberg-Richter law. The magnitudes of the observed events are plotted as points along the log-density for the neural model.}}
    \label{fig:simulated_mag_dist}
\end{figure}

Despite the synthetic catalog being complete down to $M_0 = $ 1 Mw, we only lower Mcut down to 1.7 due to the computational time it takes to find the MLE parameters of ETAS for such a large dataset. Figure \ref{fig:train_time} shows the computation time (CPU hours) to train each of the models as a function of the size of the training set using an 2.4 GHz Intel E5-2680 v4 (Broadwell) CPU. The neural model is significantly faster to train than ETAS due to the likelihood function not being dependent on the full history of the sequence, giving it complexity $O(N)$ \cite{shchur2021neural}. ETAS in contrast has complexity $O(N^2)$ due to the double sum in the likelihood.\\
Figure \ref{fig:simulated_log_scores} shows $95 \%$ confidence intervals for the log-likelihood scores on the synthetic catalogs for the varying magnitude thresholds. By varying the value of Mcut the size of the training dataset changes, depicted by the green barplots. In Figure \ref{fig:simulated_log_scores}a) the temporal log-likelihood scores for both models on the complete synthetic catalog are displayed. Although there are fluctuations, there is no significant difference between the two models' log-likelihood gain from the same Poisson baseline for all values of Mcut, suggesting the neural model has learnt to capture the ETAS data sufficiently well. Although the mean of ETAS increases as we lower Mcut, whereas the mean of the neural model fluctuates, these changes are non-significant. We speculate that we \change{don't}{do not} see significant improvement as we lower Mcut due to the fact that we are fitting temporal models to spatio-temporal data.\\
Figure \ref{fig:simulated_log_scores}b) shows the temporal log-likelihood scores for the incomplete synthetic catalog. Just as in Figure \ref{fig:simulated_log_scores}a), ETAS remains constant down to Mcut $2.0$. But now, on this incomplete dataset, the performance of ETAS significantly reduces as Mcut is lowered below this threshold. In contrast, the neural model remains constant in performance and significantly outperforms ETAS for all but the highest two thresholds, demonstrating a robustness to the missing data in this catalog. By synthetically recreating incompleteness we remove many datapoints, therefore we can fit ETAS to a lower Mcut as we \change{don't}{do not} experience the longer training times of the complete catalog.\\
Figure \ref{fig:simulated_log_scores}c) shows the magnitude log-likelihood scores for the complete synthetic catalog. For the highest two thresholds, the neural model performs significantly worse than ETAS but then remains marginally worse for all other values of Mcut. For the magnitude scores for the incomplete data in Figure \ref{fig:simulated_log_scores}d), ETAS significantly outperforms the neural model at the higher thresholds. As the threshold is lowered the two perform more similarly, owing to an increase in performance from the neural model and a slight decrease from ETAS.\\
We can understand the marginal difference in performance between the two \add{models} at lower thresholds by looking at their respective distributions. Figure \ref{fig:simulated_mag_dist} shows \change{two}{five} instances of the magnitude distribution learnt by both models at Mcut = 1.7 for the complete catalog. For ETAS we simply learn the $b$ value of the Gutenberg-Richter (GR) law whereas for the neural model, a history and time-dependent distribution for the next magnitude is learnt. In these \change{two}{five} instances, although the neural model can closely approximate the GR law, allowing it to be time-history dependent means that \change{it fluctuates around the stationary GR law}{its predictions vary across different occurrences in the sequence} and \add{therefore }in this synthetic example performs worse than the \add{stationary }data-generating distribution. Since the neural model contains orders of magnitude more parameters, this result indicates overfitting \cite{lawrence1997lessons}.
\subsection{AVN Catalog}
\begin{figure}
    \centering
    \includegraphics[width=\textwidth]{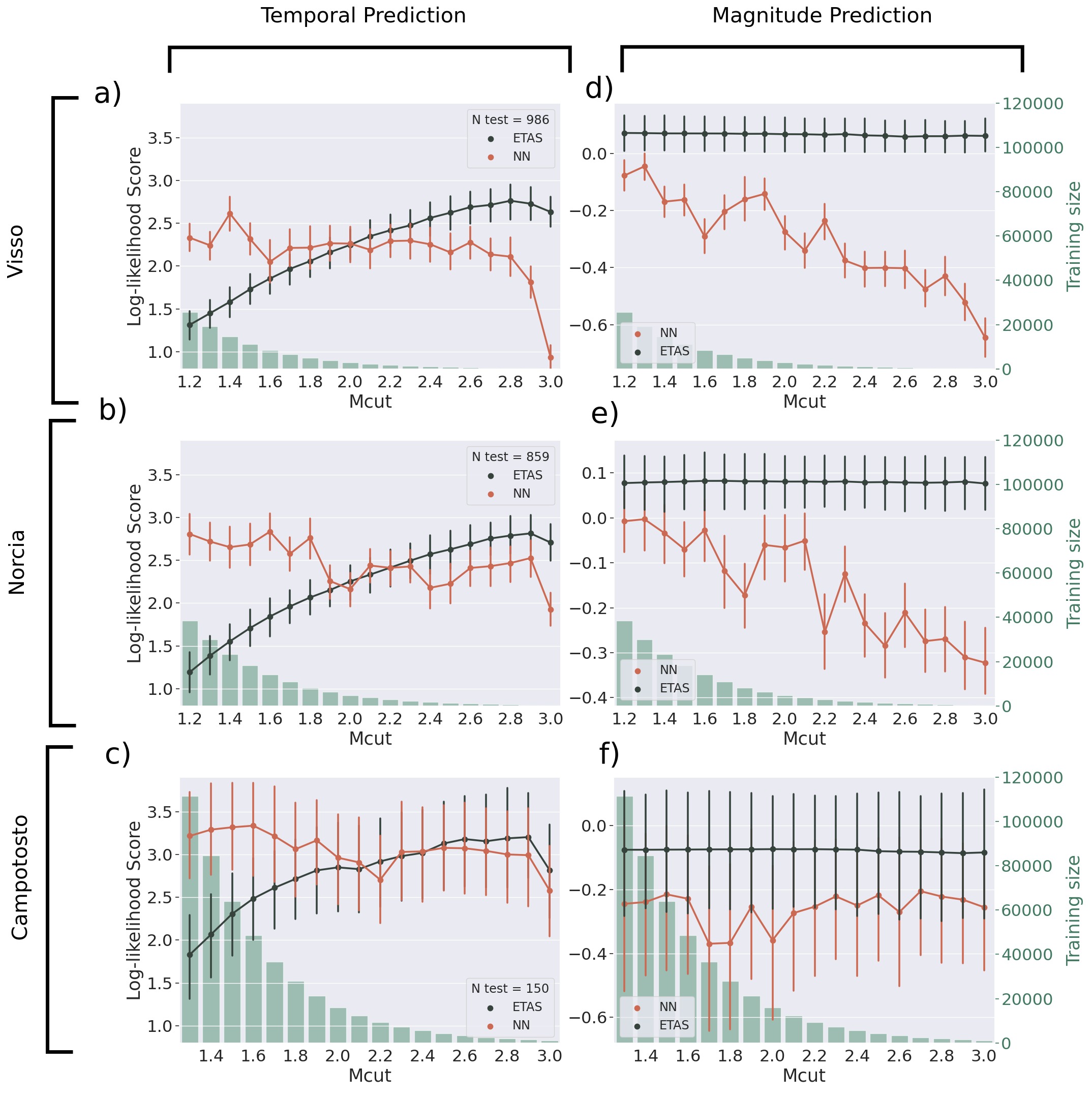}
    \caption{Results from tests on AVN catalog. 95 \% confidence intervals for the log-likelihood score of each model for varying values of Mcut. The size of the training set is displayed in the green barplot as well as the size of the testing set in the legend. a)-c) depicts the temporal log-likelihood gain from Poisson. In a), both models are trained up to the Visso earthquake, in b) both models are trained up to the Norcia earthquake and in c) both are trained up to the Campotosto earthquakes. d) - f) depict the magnitude term of the log-likelihood for the same training-testing partitions.}
    \label{fig:AVN_log_lik}
\end{figure}

\begin{figure}
    \centering
    \includegraphics[width=1.2\textwidth]{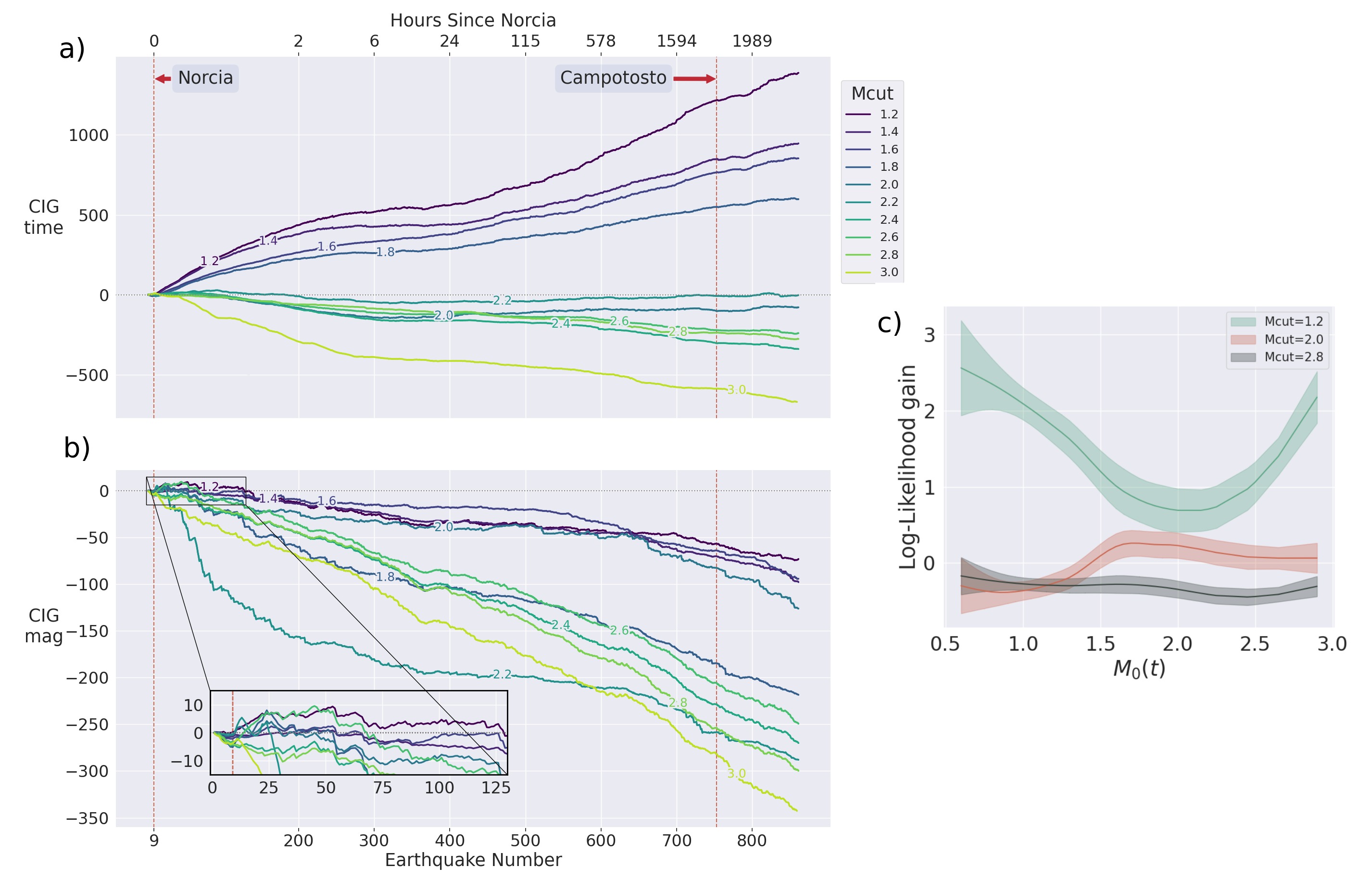}
    \caption{\add{a) - b) The Cumulative Information Gain (CIG) of the neural model over ETAS for a range of values of Mcut. The models are trained up to the Norcia earthquake and the plot depicts the evolution of the CIG from the Norcia earthquake to the end of the catalog. The curve is plotted per event, however, the actual time since the Norcia earthquake is displayed on the top axis. a) displays the CIG for event-time forecasts, b) displays the CIG for magnitude forecasts. c) displays the information gain of the neural model over ETAS as a function of the completeness of the testing catalog - both models are trained up to Norcia for Mcut = 1.2, 2.0, 2.8.}}
    \label{fig:gain_against_completeness}
\end{figure}

\begin{figure}
    \centering
    \includegraphics[width=\textwidth]{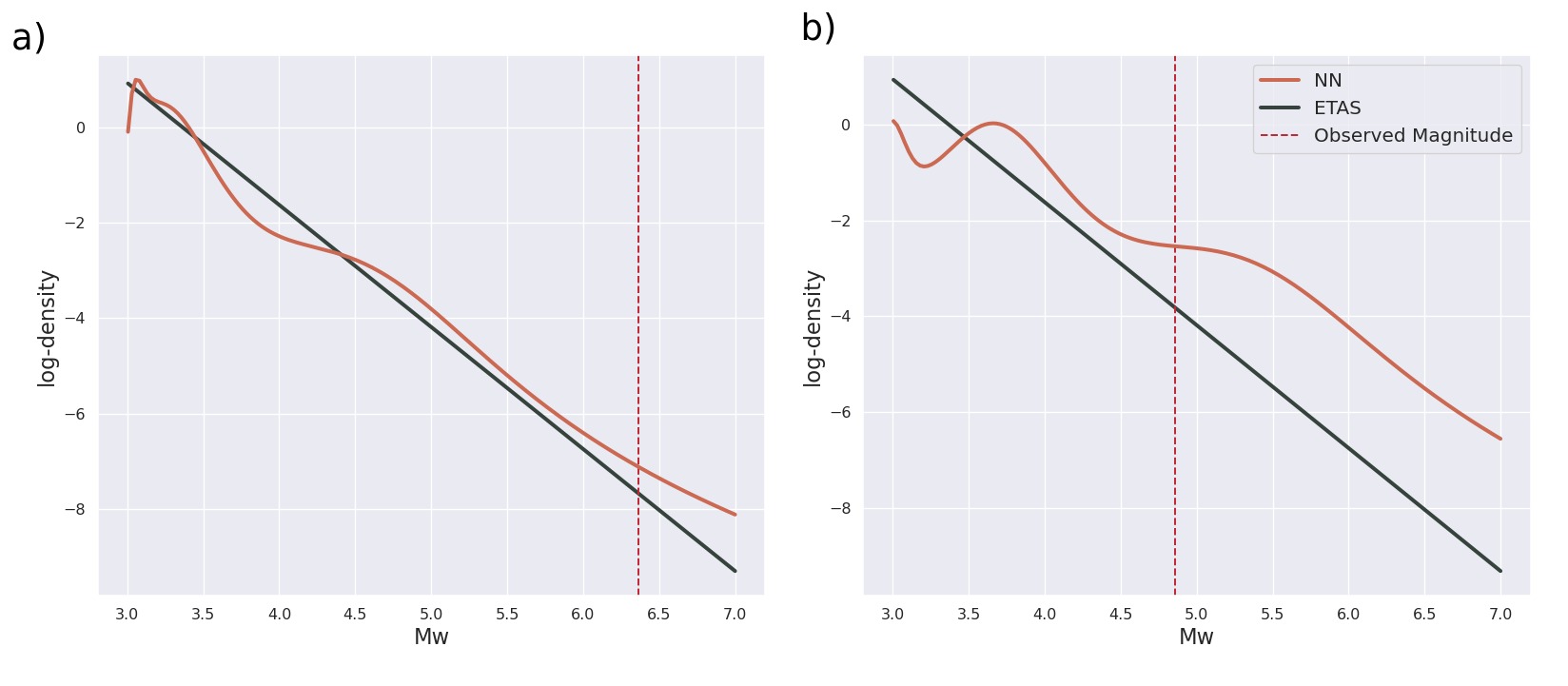}
    \caption{\add{The forecasted magnitude distribution of each model, a) at the occurrence of the Norcia earthquake, and b) at the occurrence of the next Mw3+ earthquake following Norcia.}}
    \label{fig:norcia_mag_dist}
\end{figure}

Figure \ref{fig:AVN_log_lik} shows the log-likelihood scores for both models on each of the testing-training partitions on the AVN catalog, where Figure \ref{fig:AVN_log_lik}a) is for training both models up to the Visso 5.9 Mw event, Figure \ref{fig:AVN_log_lik}b) up to the Norcia 6.5 Mw event and Figure \ref{fig:AVN_log_lik}c) up to the first of the major Campotosto earthquakes. Across all training-testing partitions Figure \ref{fig:AVN_log_lik}a)-\ref{fig:AVN_log_lik}c), as Mcut is lowered below 3 Mw, the performance of ETAS decreases consistently. The neural model, however, either remains constant in performance or improves as Mcut is lowered. In addition, as the neural model is trained on a longer period of time, its performance improves. For higher values of Mcut the neural model performs significantly worse than ETAS when trained up to Visso, but with the additional training data leading up to Norcia and Campotosto, it is similar to ETAS. For low values of Mcut, the performance of the neural model is significantly better than ETAS. When comparing across all values of Mcut\remove{, although the neural model has the highest mean gain,} neither model is significantly better than the other. Generally, the neural model is more robust to different values of Mcut than ETAS. Figure S1 of the Supporting Information shows how the fitted ETAS parameters change with Mcut.\\
The magnitude log-likelihood scores in Figure \ref{fig:AVN_log_lik}d)-\ref{fig:AVN_log_lik}f) show that the time-history dependent magnitude distribution generally cannot match the predictive power of the stationary GR law. The performance of ETAS remains constant for all values of Mcut and testing-training partitions. In Figure \ref{fig:AVN_log_lik}d) and Figure \ref{fig:AVN_log_lik}e) the neural model improves in performance as Mcut is lowered, where it only performs closely to ETAS at the very lowest threshold. This and the fact that it performs much closer to ETAS when training up to Campotosto (Figure \ref{fig:AVN_log_lik}f)) suggests that it is learning and improving when shown more data.\\
To compare the models' performance as a function of time, Figure \ref{fig:gain_against_completeness} displays the cumulative information gain (CIG) of the neural model over ETAS, for both models trained up to the Norcia earthquake. This information gain is simply the difference in the log-likelihood scores, where we subtract the score of ETAS from the neural model for both the magnitude and event-time term of the likelihood. The CIG is plotted per earthquake, but the evolution with time since the Norcia earthquake is also displayed. Figure \ref{fig:gain_against_completeness}a) shows the CIG for event time forecasts. Beyond the trend that the neural model improves over ETAS as we lower Mcut, the improvement varies over the testing catalog. For the thresholds that give the largest gain, (Mcut = 1.2, 1.4, 1.6, 1.8), the period of time with the greatest amount of gain, indicated by the steepest gradient of the curve, is found within the first 2 hours of the Norcia earthquake. This is followed by a reduced improvement up to 24 hours, beyond which it levels out and remains relatively linear.\\
Figure \ref{fig:gain_against_completeness}b) shows the CIG for the magnitude forecasts, confirming the loss in average performance of the neural model over ETAS. All thresholds decrease fairly steadily for nearly all of the testing period, ap\remove{p}art from immediately following the Norcia earthquake. For the lower thresholds the period of time following Norcia sees an improvement over ETAS very briefly before declining.\\
Figure \ref{fig:gain_against_completeness}c) shows the IG of the neural model over ETAS but now as a function of the estimate of the completeness of the testing period. Both models are trained up to Norcia for Mcut = 1.2\add{, 2.0, 2.8}. A locally weighted scatterplot smoothing (lowess) regression  \cite{cleveland1979robust} with 95 \% confidence intervals estimates this relationship. \add{For Mcut = 1.2}, the difference between the two models is smallest for intermediate values of the incompleteness (around $M_0(t) = 2.0$), but for the most complete ($M_0(t) = 1.0$) and most incomplete ($M_0(t) = 3.0$) parts of the testing catalog, the neural model performs greatest compared to ETAS. At $M_0(t) = 2.0$ where the relative performance of ETAS is best, the confidence interval for the log-likelihood difference between the two models lies above zero, centered around 0.75. So, there is no value of completeness in the testing catalog where ETAS performs as well as the neural model.\add{ For Mcut = 2.0, the neural model outperforms ETAS during more incomplete periods of the testing period and for Mcut = 2.8, ETAS is consistently better across all values of completeness. }\\
The forecasted magnitude distribution of each model shown in Figure \ref{fig:norcia_mag_dist} provides some insight into the cause of this immediate improvement over ETAS's GR forecast right after Norcia. Figure \ref{fig:norcia_mag_dist}a) shows the magnitude distribution of each model at the time of the Norcia earthquake. The two distributions resemble each other relatively closely. At the time of the next Mw3+ event, Figure \ref{fig:norcia_mag_dist}b), the neural model has shifted its density towards higher magnitude values\add{ anticipating a lack of observed smaller magnitude earthquakes due to catalog incompleteness}.
\section{Discussion}
\subsection{Approximating ETAS}\label{Aprox_ETAS}
The ability to approximate ETAS data using a neural point process is a benchmark goal. Demonstrating a baseline level of expressiveness is essential before any work on real data is done. Given other neural point process models regularly use univariate or discretely marked Hawkes data as a baseline \cite{du2016recurrent,NEURIPS2019_39e4973b}, this result provides an example of fit to continuous bivariate Hawkes data.\\ 
Specifically, the merit of this fit to ETAS data is that it uses a truncated history of events. Truncating the ETAS model to sum over only the last $d$ events would dramatically change the evaluation of the intensity function between a significantly large earthquake $d$ events ago or $d+1$ events ago, thus instead the way in which to formulate the relationship between the intensity and these truncated events is learnt. The past $d$ events are exhibiting behaviour based on the events prior and thus we \change{don't}{do not} directly specify the contribution from events further back in time, instead dependence on such events is learnt indirectly.\\
The reduction in the amount of history each forecast is dependent upon drastically improves the computational requirements for both learning and evaluation of the likelihood. To create forecasting models alongside the growing size of earthquake catalogs generated through machine learning based phase picking, we require models that scale well with the data. Shown here is that we can achieve the same predictive accuracy as ETAS (in a synthetic catalog) but with a smaller computational budget.

\subsection{Embracing and Ignoring Data Incompleteness}

At the previously explored thresholds of this sequence, \cite{ebrahimian2017robust,marzocchi2017earthquake,mancini2019improving, mancini2022use}, the neural point process performs similarly to ETAS. The biggest deviations between the two models are found as the magnitude threshold is lowered into new unexplored regions revealed by this enhanced catalog. The deviations come from the fact that the neural model increases gradually in performance as the threshold is lowered whereas ETAS drastically decreases in performance. Below, we offer an interpretation for these results. \\
We argue that the largest gains made by the neural model are due to its ability to handle the incomplete data immediately following large earthquakes. There are two justifications for this: Given that the magnitude threshold of the input catalog is lowered into regions where there are periods when $M_0(t) > \text{Mcut} $, we expect there to be biases in fitting ETAS \cite{zhuang2017data, seif2017estimating,hainzl2016rate}. The consequences of these biases on ETAS are reflected in the log-likelihood scores on the synthetic incomplete data figure \ref{fig:simulated_log_scores}b). In this synthetic catalog with short-term aftershock incompleteness, ETAS drastically reduces in performance in contrast to the neural model. Since a basic ETAS is formulated as completely observing all potentially triggering earthquakes it poorly captures incomplete sequences. The same shape of plot is found in the real data in Figure \ref{fig:AVN_log_lik}a)-c), where the performance of ETAS decreases as Mcut is lowered. In contrast, other studies have found decreasing the minimum triggering magnitude improves the performance of ETAS (e.g., \citeA{werner2011high, helmstetter2014adaptive}). A consequence of the bias in the ETAS parameters is that the forecasting performance is only competitive with the neural model during intermediate values of incompleteness, Figure \ref{fig:gain_against_completeness}c). Even during complete periods in the testing catalog, since ETAS has been fit on incomplete data, it fails to forecast well. \\
The second justification comes through considering the process by which the observed data are generated, e.g. as described by \citeA{omi2014estimating}. The relationship between the underlying process $\lambda(t|H_t)$ and the observed process $v(t|H_t)$ that forms the catalog itself can be written as
\begin{linenomath*}
\begin{equation*}
    v(t|H_t) = \lambda(t|H_t)\ r(t|M_0),
\end{equation*}
\end{linenomath*}
where $r(t|M_0)$ is the probability of detection at time t. For ETAS variants that deal with time-varying completeness \cite<e.g.>{omi2014estimating,hainzl2016apparent,hainzl2016rate,mizrahi2021embracing}, this function has to be estimated alongside the parameters of ETAS.
When fitting a temporal ETAS model to data with temporal incompleteness, bias in the fitted parameters comes from the modeling assumption that $v^*(t) = \lambda^*(t)$. Through the use of a flexible model such as this neural point process, rather than trying to learn both the underlying process and the detection rate, we instead directly learn the observed process. So in fact, in the construction of the model rather than equation (\ref{eq:NN_modeling_choice}), the observation process is approximated,
\change{\protect\begin{linenomath*}
\begin{equation*}
    \phi(t-t_i,\mathbf{h}_i) = v^*(t),
\end{equation*}
\end{linenomath*}}{\protect\begin{linenomath*}
\begin{equation*}
    \phi(t-t_i|\mathbf{h}_i) = v^*(t),
\end{equation*}
\end{linenomath*}}
where $t_i$ is the time of the last event. As we lower Mcut into regions of temporal incompleteness, unlike ETAS, the performance of the neural model does not decrease, Figure \ref{fig:simulated_log_scores}b). This demonstrates the neural models' ability to fit to observed data as it is not biased by an increasing amount of missingness.\\
Learning to model the observation process requires the assumption that the process of the incompleteness is stationary for future forecasts, thus if there is new methodology in data collection, the model would have to be re-trained on this new data. This is similar to detection rate based methods, \cite{omi2014estimating,hainzl2016apparent,hainzl2016rate,mizrahi2021embracing}, that would also need to update their detection function with new observational methodology.\\
To further test the effectiveness of the neural model, a comparison with other ETAS models that specifically deal with incompleteness is needed. But given that methods that deal with incompleteness only increase the computational requirements upon fitting a basic ETAS model, neural point processes could offer a more efficient way to deal with missing data. This is especially important when moving towards using enhanced catalogs such as the one used here. Temporal variations in completeness must be considered when using these catalogs and to be able to use these catalogs we must take more care with the computational efficiency of models.\\
Although data incompleteness is the most reasonable argument for the significant gains of the neural point process at low magnitude thresholds, we shouldn't limit ourselves to this description. We have reached this conclusion by extrapolating the forecasting results on synthetic data and through arguments about which statistical process is being approximated by the neural model. However, we shouldn't rule out the possibility that the new low magnitude data in this enhanced catalog has contributed additional signal that is not explainable by ETAS. Even for the intermediate value of completeness that results in the best performance of ETAS compared with the neural model, there is still an average log-likelihood gain of around 0.75, Figure \ref{fig:gain_against_completeness}c).\add{ We can compare this with Figure S3c) in the supporting information, where on the incomplete synthetic data we see a truncated curve of the same shape. In this synthetic experiment, there are periods of completeness where the performance between the two models is comparable, however on the real data,} \change{S}{s}ince there is no value of completeness \change{in the testing catalog where ETAS performs as well as the neural model, we speculate}{ that results in comparable performance, this would suggest} that something additional is contributing to the gains. \change{This}{The} question of whether there is additional signal in low magnitude events found in enhanced catalogs such as this one needs further attention beyond this study. We believe that further development in neural point processes will aid modellers in analysing this wealth of new data as neural models provide more flexible modelling alternatives and can cope with the scale of new enhanced catalogs.
\subsection{Limitations}
This study presents a flexible model that does not suffer from the same misspecification as ETAS due to short-term aftershock incompleteness. Although the size of the gains for these magnitude thresholds is large, we found, however, no significant overall improvement in forecasting ability over ETAS across magnitude thresholds. Comparing the value of Mcut that gives the greatest performance for each model finds that although the mean of the neural model is highest, the gain over ETAS is not significant. In this two dimensional time-magnitude domain, given the flexibility of this neural network and the data volume provided, \remove{we hypothesise that} there is insufficient signal in the data to learn anything significantly better than ETAS.\add{ This would suggest that the time and magnitude data from low magnitude events alone does not give us additional information in forecasting M3+ events.} This motivates considering whether additional features can aid in the forecasting ability of neural point processes. Given that operationally we also require spatial forecasts, this is an obvious future extension to the model. It is natural to expect that including spatial covariates would improve forecasting performance \cite{utsu1955relation,ogata1998space}, however, it is not clear that considering them as an additional dimension to the input of an RNN would learn any spatial structure from the data. Neural point processes for spatio-temporal data do not utilise RNNs which are primarily sequence encoders and instead consider models based on Ordinary Differential Equations (ODEs) \cite{chen2020neural,bilovs2021neural}. We believe such models should outperform RNN-based models on spatio-temporal data.\\
By modelling the magnitudes by a completely unconstrained density that is also time-history dependent, we create lots of potential for over-fitting to the data \cite{ying2019overview}. This is exactly what is observed during the tests on synthetic data where we learn a `noisy' Gutenberg-Richter law. It is the likely source for the performance which is on average worse than ETAS on the AVN catalog. However, by letting this function be unconstrained, the model was able to make improvements over ETAS immediately following Norcia. This isn't too surprising since deviations from a stationary GR law have been observed, either through fluctuations of the b-value in space or time \cite{schorlemmer2005variations, gulia2018effect} or short-term aftershock incompleteness \cite{kagan2004short,woessner2005assessing,helmstetter2006comparison}.\\
A final limitation of the model presented here is its (in)ability to simulate events into the future. Where simulation of ETAS can be done due its equivalent branching process formulation, simulation from the neural model can only be leveraged through the intensity function at a given point in time with a thinning algorithm \cite{ogata1981lewis}. But given that calculation of the intensity with a neural point process is much more efficient than ETAS, it would be quicker to do thinning simulation with a the neural model. Slower simulation can be overcome by using an alternative flexible formulation of the likelihood such as the one used in \citeA{shchur2019intensity}, however, we would lack the ability to target earthquakes above a magnitude threshold.
\section{Conclusion}
We present an initial investigation into the viability of neural point processes for the forecasting of short term seismicity. The neural point process is formulated in a similar way to the ETAS model, only with a much more flexible way of representing the intensity function. Now with much larger earthquake catalogs, data-driven point process models present us with an opportunity to investigate whether these new data may offer some deviation to the parameterization of ETAS as well as providing more computationally efficient models that are robust to missing data.\\
We extend the existing point process model of \citeA{NEURIPS2019_39e4973b} so that it also models the magnitudes associated with the events contained in earthquake sequences. We also show how this model can be used to forecast earthquakes above some target threshold magnitude through decomposing the cumulative hazard function between target events. A notable feature of the presented model is that a forecast is only dependent on a fixed length vector representing the history of events, making the evaluation of the likelihood scale linear with the sample size.\\
With an experiment on data simulated from the ETAS model we demonstrate this computational advantage by showing a stark improvement on the time to train the neural point process against the ETAS model, whilst still obtaining a similar likelihood score on test data. We find that defining a more flexible time-history dependent magnitude distribution leads to overfitting and consequentially the magnitude likelihood scores are worse than when using a stationary Gutenberg-Richter law.\\
Through artificially removing events from the synthetic catalog we create a dataset that mimics short-term aftershock incompleteness. We find that on this altered catalog, the performance of ETAS now decreases as the magnitude threshold is lowered. In contrast, the neural model remains constant in performance, suggesting that it is more robust to the missing data found in typical earthquake catalogs.\\
On real data from the Amatrice-Visso-Norcia sequence the performance of both models vary with respect to the magnitude threshold of the input catalog. Both models perform similarly at previously explored thresholds (Mw3+), but when lowered into magnitude regions revealed by the new catalog, ETAS decreases in performance unlike the neural point process. We argue that these gains are due to the neural model's ability to handle the incomplete data found in this enhanced catalog. This experiment both motivates the need for considering temporal completeness when using enhanced catalogs and motivates further work into what the spatial covariates of this dataset might offer when combined with flexible point process models such as this one.

\section{Open Research}
The Amatrice-Visso-Norcia catalog produced by \citeA{tan2021machine} is accessible at the Zenodo repository https://doi.org/10.5281/zenodo.4736089 \cite{https://doi.org/10.5281/zenodo.4736089}. The ETAS simulator used to generate the synthetic data was written for \citeA{mizrahi2021effect} and \citeA{mizrahi2021embracing}, and is available at \url{https://github.com/lmizrahi/etas}, \cite{mizrahiETAS}. Both synthetic and real datasets are found in the reproducibility package along with the models and the experimental design used in this study \cite{Neural-Point-Process}.







\acknowledgments
This project is funded by Compass - Centre for Doctoral Training in Computational Statistics and Data Science (EPSRC Grant Ref EP/S023569/1). Compass is funded by United Kingdom Research and Innovation (UKRI) through the Engineering and Physical Sciences Research Council (EPSRC), \url{https://www.ukri.org/councils/epsrc}. This project also has received funding from the European Research Council (ERC) under the European Union’s Horizon 2020 research and innovation programme (grant agreement no. 821115, Real-time earthquake rIsk reduction for a reSilient Europe (RISE), \url{http://www.rise-eu.org}).


%
%



\bibliography{agusample.bib}

%
%
%
%
%

\end{document}


%
%


\title{Supporting Information for ``Forecasting the 2016-2017 Central Apennines Earthquake Sequence with a Neural Point Process"}
%
%

%
%



\authors{Samuel Stockman \affil{1}, Daniel J. Lawson \affil{1}, Maximilian J. Werner \affil{2}}


\affiliation{1}{School of Mathematics, University of Bristol}
\affiliation{2}{School of Earth Sciences, University of Bristol}

\begin{article}

\noindent\textbf{Contents of this file}
\begin{enumerate}
\item Figures S1 to S3 

\end{enumerate}

\noindent\textbf{Introduction} \\
In this supplement, we share the learnt ETAS parameters for the Amatrice-Visso-Norcia catalog for all the values of Mcut and all training testing partitions (Figure S1). Furthermore we show the cummulative information gain (CIG) of the neural model over ETAS on the complete synthetic catalog for both time and magnitude forecasting (Figure S2), as well as for the incomplete synthetic catalog (Figure S3).

\end{article}
\clearpage

\begin{figure}
    \centering
    \includegraphics[width=0.7\textwidth]{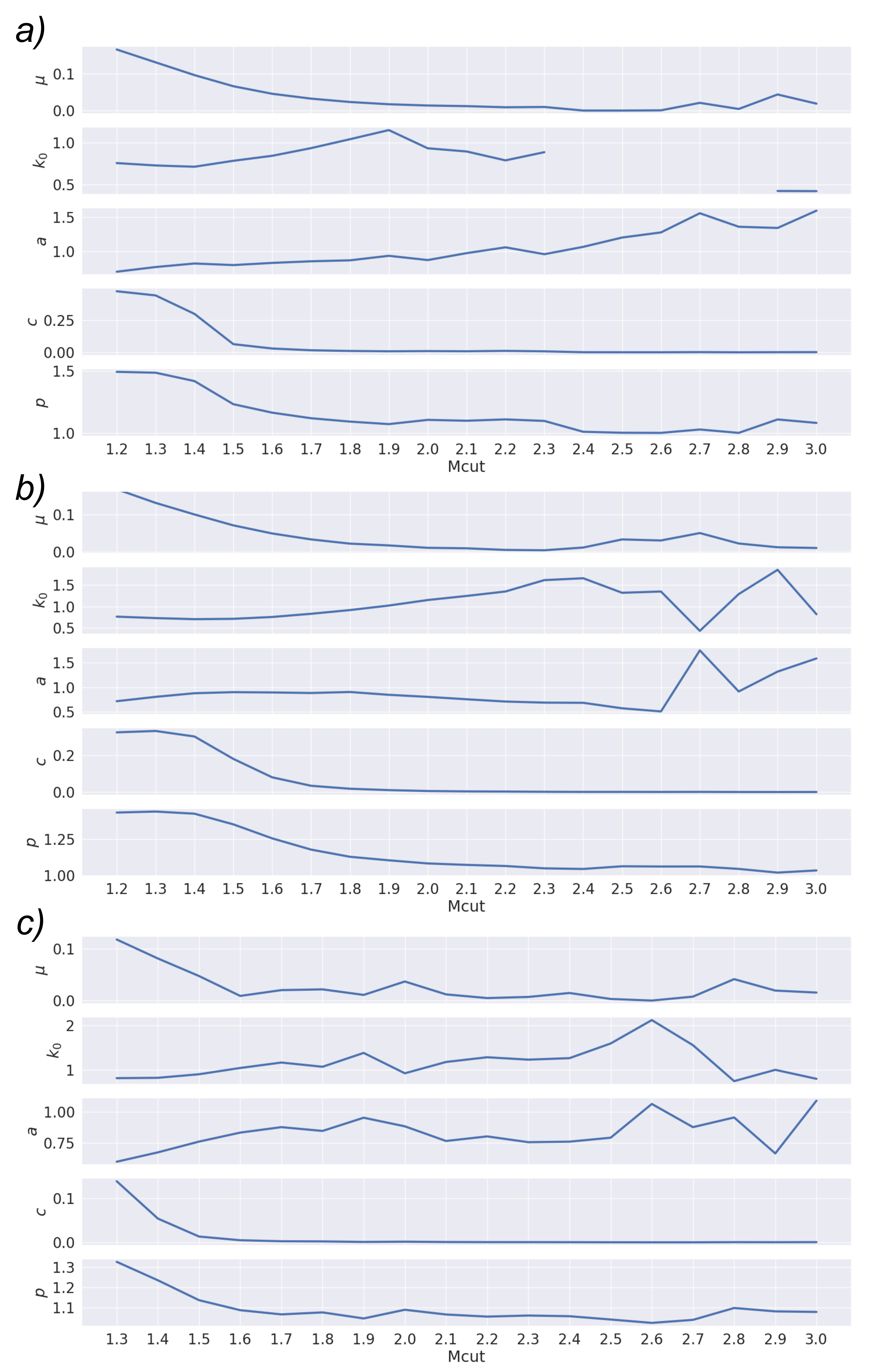}
    \caption{Fitted ETAS parameters as a function of Mcut. \textbf{a)} training up to the Visso earthquake. $k_0$ parameters for Mcut 2.4-2.8 have been removed from the plot to aid in visualisation. These parameter values are orders of magnitude larger. \textbf{b)} training up to the Norcia earthquake. \textbf{c)} training up to the Campotosto earthquakes. The unit of time is hours.}
    \label{fig:params}
\end{figure}

\begin{figure}
    \centering
    \includegraphics[width=\textwidth]{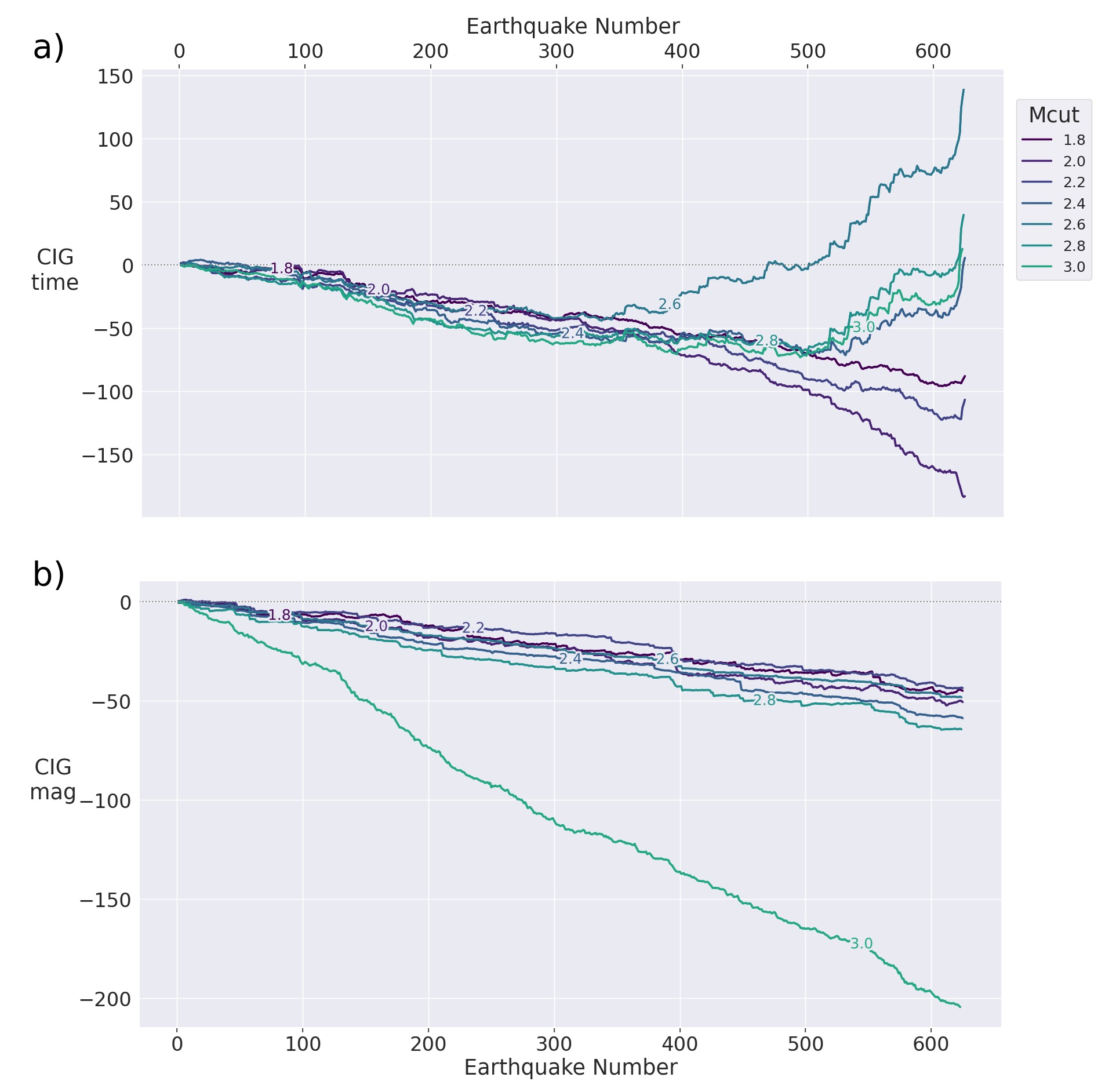}
    \caption{a) - b) The Cumulative Information Gain (CIG) of the neural model over
    ETAS for a range of values of Mcut. The models are trained and forecasted on the complete synthetic catalog
    and the plot depicts the evolution of the CIG from the beginning of the testing period to the end of
    the catalog. a) displays the CIG for event-time forecasts, b)
    displays the CIG for magnitude forecasts.}
    \label{fig:params}
\end{figure}

\begin{figure}
    \centering
    \includegraphics[width=\textwidth]{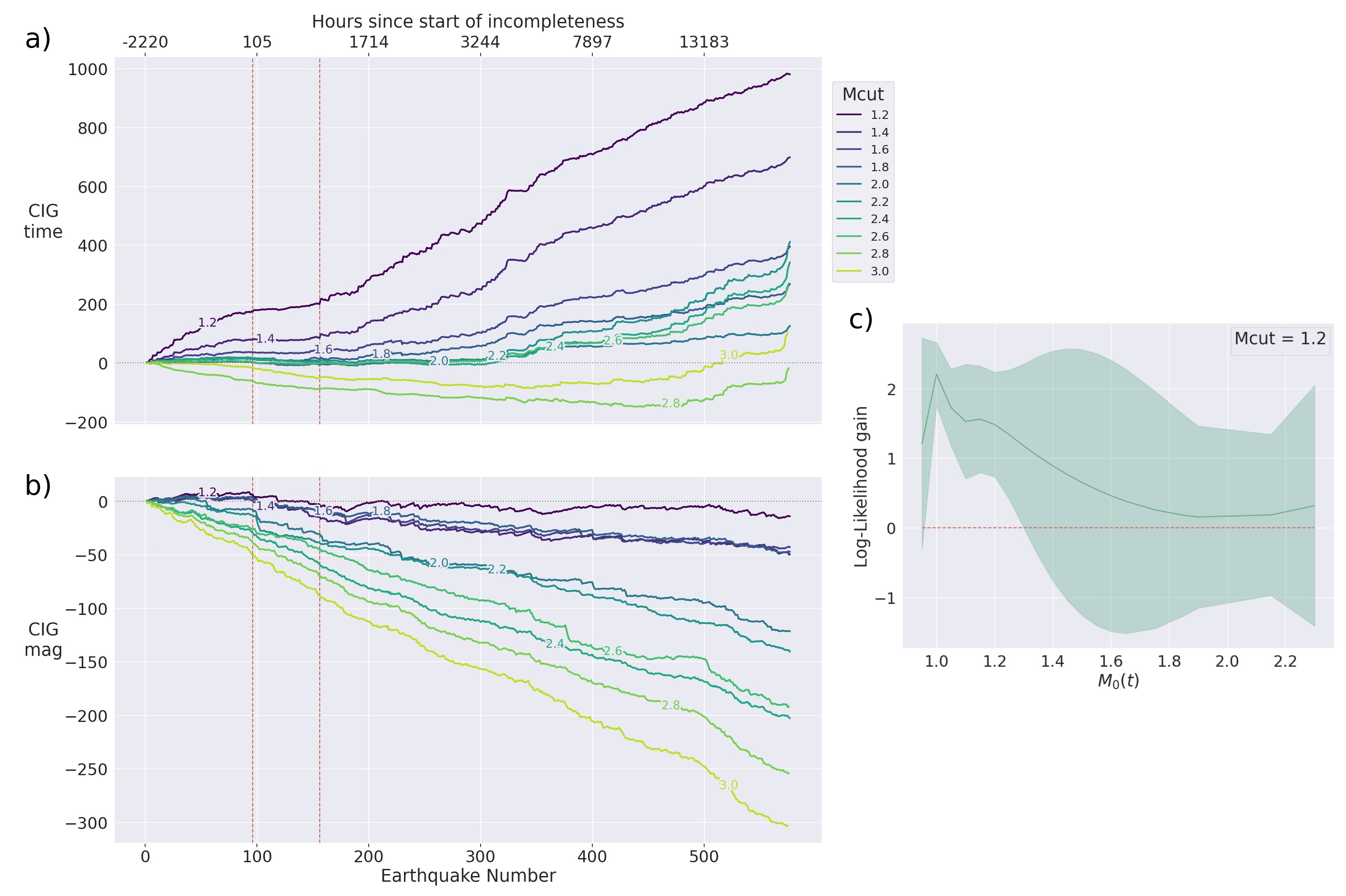}
    \caption{a) - b) The Cumulative Information Gain (CIG) of the neural model over
    ETAS for a range of values of Mcut. The models are trained and forecasted on the incomplete synthetic catalog
    and the plot depicts the evolution of the CIG from the beginning of the testing period to the end of
    the catalog. The curve is plotted per event, however, the time since the start of a period of incompleteness is displayed on the top axis. a) displays the CIG for event-time forecasts, b)
    displays the CIG for magnitude forecasts. c) displays the information gain of the neural
    model over ETAS as a function of the completeness of the testing catalog - both models
    are trained with Mcut = 1.2.}
    \label{fig:params}
\end{figure}